\documentclass[12pt]{article}
\usepackage{amsmath,amssymb}
\usepackage{iftex}
\ifPDFTeX
  \usepackage[T1]{fontenc}
  \usepackage[utf8]{inputenc}
  \usepackage{textcomp} % provide euro and other symbols
\else % if luatex or xetex
  \usepackage{unicode-math}
  \defaultfontfeatures{Scale=MatchLowercase}
  \defaultfontfeatures[\rmfamily]{Ligatures=TeX,Scale=1}
\fi
\usepackage{lmodern}
\usepackage{enumitem}
\ifPDFTeX\else  
\fi
\IfFileExists{upquote.sty}{\usepackage{upquote}}{}
\IfFileExists{microtype.sty}{% use microtype if available
  \usepackage[]{microtype}
  \UseMicrotypeSet[protrusion]{basicmath} % disable protrusion for tt fonts
}{}
\makeatletter
\@ifundefined{KOMAClassName}{% if non-KOMA class
  \IfFileExists{parskip.sty}{%
    \usepackage{parskip}
  }{% else
    \setlength{\parindent}{0pt}
    \setlength{\parskip}{6pt plus 2pt minus 1pt}}
}{% if KOMA class
  \KOMAoptions{parskip=half}}
\makeatother
\usepackage[table,dvipsnames]{xcolor}
\makeatletter
\ifx\paragraph\undefined\else
  \let\oldparagraph\paragraph
  \renewcommand{\paragraph}{
    \@ifstar
      \xxxParagraphStar
      \xxxParagraphNoStar
  }
  \newcommand{\xxxParagraphStar}[1]{\oldparagraph*{#1}\mbox{}}
  \newcommand{\xxxParagraphNoStar}[1]{\oldparagraph{#1}\mbox{}}
\fi
\ifx\subparagraph\undefined\else
  \let\oldsubparagraph\subparagraph
  \renewcommand{\subparagraph}{
    \@ifstar
      \xxxSubParagraphStar
      \xxxSubParagraphNoStar
  }
  \newcommand{\xxxSubParagraphStar}[1]{\oldsubparagraph*{#1}\mbox{}}
  \newcommand{\xxxSubParagraphNoStar}[1]{\oldsubparagraph{#1}\mbox{}}
\fi
\makeatother

\usepackage{longtable,booktabs,array}
\usepackage{calc} % for calculating minipage widths
\usepackage{etoolbox}
\makeatletter
\patchcmd\longtable{\par}{\if@noskipsec\mbox{}\fi\par}{}{}
\makeatother
\IfFileExists{footnotehyper.sty}{\usepackage{footnotehyper}}{\usepackage{footnote}}
\makesavenoteenv{longtable}
\usepackage{graphicx}
\makeatletter
\def\maxwidth{\ifdim\Gin@nat@width>\linewidth\linewidth\else\Gin@nat@width\fi}
\def\maxheight{\ifdim\Gin@nat@height>\textheight\textheight\else\Gin@nat@height\fi}
\makeatother
\setkeys{Gin}{width=\maxwidth,height=\maxheight,keepaspectratio}
\makeatletter
\def\fps@figure{htbp}
\makeatother

\makeatletter
\@ifpackageloaded{caption}{}{\usepackage{caption}}
\AtBeginDocument{%
\ifdefined\contentsname
  \renewcommand*\contentsname{Table of contents}
\else
  \newcommand\contentsname{Table of contents}
\fi
\ifdefined\listfigurename
  \renewcommand*\listfigurename{List of Figures}
\else
  \newcommand\listfigurename{List of Figures}
\fi
\ifdefined\listtablename
  \renewcommand*\listtablename{List of Tables}
\else
  \newcommand\listtablename{List of Tables}
\fi
\ifdefined\figurename
  \renewcommand*\figurename{Figure}
\else
  \newcommand\figurename{Figure}
\fi
\ifdefined\tablename
  \renewcommand*\tablename{Table}
\else
  \newcommand\tablename{Table}
\fi
}
\@ifpackageloaded{float}{}{\usepackage{float}}
\floatstyle{ruled}
\@ifundefined{c@chapter}{\newfloat{codelisting}{h}{lop}}{\newfloat{codelisting}{h}{lop}[chapter]}
\floatname{codelisting}{Listing}

\makeatother
\makeatletter
\@ifpackageloaded{caption}{}{\usepackage{caption}}
\@ifpackageloaded{subcaption}{}{\usepackage{subcaption}}
\makeatother

\ifLuaTeX
  \usepackage{selnolig}  % disable illegal ligatures
\fi
\usepackage[]{natbib}
\usepackage{bookmark}

\IfFileExists{xurl.sty}{\usepackage{xurl}}{} % add URL line breaks if available
\hypersetup{
  pdftitle={Title},
  pdfauthor={Author 1; Author 2},
  pdfkeywords={3 to 6 keywords, that do not appear in the title},
  colorlinks=true,
  linkcolor={blue},
  filecolor={Maroon},
  citecolor={Blue},
  urlcolor={Blue},
  pdfcreator={LaTeX via pandoc}}

\usepackage{mathtools}
\usepackage{amsthm}
\usepackage{bm}
\usepackage{amsfonts}
\usepackage{dsfont}
\usepackage{bbold}
\usepackage{tabularx}
\usepackage{makecell}
\usepackage{multirow}
\usepackage{arydshln}
\usepackage{lscape}
\usepackage{rotating}
\usepackage[flushleft]{threeparttable}
\usepackage{siunitx}
\usepackage{placeins}
\usepackage{afterpage}
\usepackage{algorithm}
\usepackage{algpseudocode}
\usepackage[scr]{rsfso}
\usepackage{pgffor}
\usepackage{comment}
\usepackage{setspace}
\usepackage[normalem]{ulem}
\usepackage{adjustbox}
\usepackage{pgfplots}
 \pgfplotsset{compat=1.18}
\usepackage{tikz}
\usepackage{etoc}
\DeclareMathOperator*{\argmin}{arg\,min}

\newcommand{\norm}[1]{{\left\lVert#1\right\rVert}}

\definecolor{applegreen}{rgb}{0.55, 0.71, 0.0}
\definecolor{ao(english)}{rgb}{0.0, 0.5, 0.0}

\newcommand{\bA}{\bm A}

\newcommand{\bD}{\bm D}

\newcommand{\be}{\bm e}

\newcommand{\bg}{\bm g}

\newcommand{\bh}{\bm h}
\newcommand{\bI}{\bm I}

\newcommand{\bM}{\bm M}

\newcommand{\br}{\bm r}

\newcommand{\bs}{\bm s}

\newcommand{\bu}{\bm u}
\newcommand{\bV}{\bm V}
\newcommand{\bv}{\bm v}

\newcommand{\bX}{\bm X}
\newcommand{\bx}{\bm x}

\newcommand{\by}{\bm y}

\newcommand{\bzero}{\bm 0}

\newcommand{\bbeta}{\bm \beta}

\newcommand{\bdelta}{\bm \delta}

\newcommand{\bnu}{\bm \nu}

\newcommand{\bchi}{\bm \chi}

\newcommand{\bGamma}{\bm \varGamma}
\newcommand{\bDelta}{\bm \varDelta}

\newcommand{\bSigma}{\bm \varSigma}

\newcommand{\supp}{\mbox{supp}}

\theoremstyle{definition}

\newtheorem{assumption}{Assumption}
\newtheorem{example}{Example}
\newtheorem{remark}{Remark}

\theoremstyle{plain}
\newtheorem{proposition}{Proposition}
\newtheorem{theorem}{Theorem}
\newtheorem{lemma}{Lemma}
\newtheorem{corollary}{Corollary}

\makeatletter
\def\adl@drawiv#1#2#3{%
        \hskip.5\tabcolsep
        \xleaders#3{#2.5\@tempdimb #1{1}#2.5\@tempdimb}%
                #2\z@ plus1fil minus1fil\relax
        \hskip.5\tabcolsep}
\newcommand{\cdashlinelr}[1]{%
  \noalign{\vskip\aboverulesep
           \global\let\@dashdrawstore\adl@draw
           \global\let\adl@draw\adl@drawiv}
  \cdashline{#1}
  \noalign{\global\let\adl@draw\@dashdrawstore
           \vskip\belowrulesep}}
\makeatother



\usepackage{footmisc}
\usepackage{hyperref}
\usepackage{setspace}
\newcommand{\anon}{1}

\allowdisplaybreaks[3]
\begin{document}

\def\spacingset#1{\renewcommand{\baselinestretch}%
{#1}\small\normalsize} \spacingset{1}

\makeatletter
\g@addto@macro\normalsize{%
  \setlength{\abovedisplayskip}{6pt plus 2pt minus 2pt}%
  \setlength{\belowdisplayskip}{6pt plus 2pt minus 2pt}%
  \setlength{\abovedisplayshortskip}{3pt plus 2pt minus 1pt}%
  \setlength{\belowdisplayshortskip}{4pt plus 2pt minus 1pt}%
}
\makeatother
%%%%%%%%%%%%%%%%%%%%%%%%%%%%%%%%%%%%%%%%%%%%%%%%%%%%%%%%%%%%%%%%%%%%%%%%%%%%%%

\if1\anon
{
  \title{\bf Broken Adaptive Ridge in High-Dimensional Vector Autoregressions}
  \author{Gianluca Cubadda\hspace{.2cm}\\
  University of Rome Tor-Vergata \vspace{.5cm}\\
  Luca Margaritella\hspace{.2cm}\\
    Lund University\vspace{.5cm}\\
    Orest Prifti\thanks{The views expressed in this article are the author's only and not those of the Bank of Italy.} \\
    Bank of Italy,\\University of Rome Tor-Vergata}
  \maketitle
} \fi

\if0\anon
{
  \bigskip
  \bigskip
  \bigskip
  \begin{center}
    {\LARGE\bf Broken Adaptive Ridge in High-Dimensional Vector Autoregressions}
\end{center}
  \medskip
} \fi

\bigskip
\begin{abstract}
We study regularized estimation in high-dimensional vector autoregressions under physical dependence and row-wise approximate sparsity using the Broken Adaptive Ridge (BAR) estimator. BAR is an iteratively reweighted Ridge procedure whose fixed-point first-order conditions imply a coordinatewise thresholding rule that produces exact zeros. We show that the BAR estimation error can be analyzed on the union of the population support above the BAR threshold and the support selected in the sample. The size of this union can be controlled at the rate implied by approximate sparsity, allowing us to establish the required restricted-curvature condition and derive $\ell_1$, $\ell_2$, and prediction error rates when the number of regressors grows polynomially with the sample size and may exceed it. Under exact sparsity, the resulting rates recover the familiar rates for sparse high-dimensional estimation. Monte Carlo experiments across a range of high-dimensional VAR designs show that BAR delivers strong estimation and forecasting performance, and retains a high share of relevant coefficients relative to Lasso, adaptive Lasso, and SCAD. An application to U.S. inflation forecasting using FRED-MD shows particularly strong performance at the one-month horizon.
\end{abstract}

\noindent%
\vfill

% Table of content in the main text
\etocdepthtag.toc{main}

\begingroup
\etocsettagdepth{main}{subsubsection}     % main text
\etocsettagdepth{appendix}{section}    % appendix
\tableofcontents
\endgroup

\newpage
\spacingset{1.8} % DON'T change the spacing!

\section{Introduction}
The Lasso \citep{tibshirani1996regression} solves a convex relaxation of the ideal but computationally infeasible $\ell_0$-penalized least squares problem. This ``convexification'' yields tractability, but the linear $\ell_1$ penalty also shrinks large coefficients, potentially introducing substantial bias on strong signals. Following its introduction, a large literature has developed methods aimed at reducing this shrinkage bias, including the adaptive Lasso of \citet{zou2006adaptive} and folded-concave penalties such as SCAD \citep{fan2001variable}. These approaches modify the uniform penalization induced by the Lasso, either through data-dependent weights or nonconvex penalty functions, so that sufficiently large coefficients are penalized less severely.

Broken Adaptive Ridge (BAR) \citep{dai2018broken} belongs broadly to this family of adaptive regularization methods, but with a different geometry. Instead of a weighted $\ell_1$ penalty, BAR uses an adaptive quadratic penalty implemented through iteratively reweighted Ridge steps. Each reweighting step solves a strictly convex weighted-Ridge problem, and the resulting estimates determine the penalty weights used in the next iteration. The limiting estimator also inherits a Ridge-type grouping effect under strong collinearity. At the same time, the penalty curvature is highly heterogeneous: small coefficients face very steep curvature and are pushed toward zero, whereas large coefficients face almost no shrinkage. In this sense, BAR acts as a closer surrogate to the ideal $\ell_0$ penalty, which primarily cares about whether a coefficient is active and leaves strong signals essentially unshrunk.

These features are particularly appealing in high-dimensional vector autoregressions (VARs), where strong collinearity across variables and across lags is the rule rather than the exception. In such settings, regularization should primarily decide which lagged effects enter the model, without unduly shrinking the dominant autoregressive coefficients. BAR is well suited to this task: strong own-lags are kept nearly unbiased, weak cross-lags are aggressively shrunk toward zero, and highly correlated lagged regressors are treated in a more grouped fashion. By contrast, even adaptive $\ell_1$ procedures remain tied to an $\ell_1$ geometry, which can still induce substantial shrinkage bias on dominant autoregressive effects.

Regularization of large VARs has been developed primarily around $\ell_1$-based methods. \citet{basu2015regularized} establish high-dimensional estimation bounds for sparse VARs under temporal dependence, while \citet{kock2015oracle} derive oracle inequalities for the Lasso and adaptive Lasso. \citet{nicholson2020high} develop hierarchical group-Lasso penalties for high-dimensional VAR forecasting, allowing the lag structure itself to be regularized. Related sparsity concepts for high-dimensional VARs are discussed by \citet{krampe2021sparsity}. BAR itself was introduced by \citet{dai2018broken}, with the scalable coordinatewise cycBAR implementation developed by \citet{kawaguchi2021scalable}. BAR has also been considered by \citet{huang2025model}, specifically for model selection in increasing-order sparse VARs under full-rank design conditions, a substantially different setting compared to ours.

This paper develops the theory of BAR estimation for high-dimensional VARs, allowing for both temporal and cross-sectional dependence. Under physical dependence of the underlying process and row-wise weak sparsity of the VAR coefficient matrices, we derive high-probability prediction and $\ell_1/\ell_2$ estimation error bounds under polynomial dimensional growth, including regimes in which the number of regressors exceeds the sample size. The analysis exploits a feature specific to BAR: its fixed-point thresholding rule allows the estimation error to be analyzed on the union of coefficients above the BAR threshold in the population and those selected by BAR in the sample. We show that the size of this union can be controlled at the relevant sparsity rate, after which restricted curvature is established on this set. This differs from standard $\ell_1$-based approaches, where curvature is typically imposed over a cone of directions. Weak sparsity allows many small nonzero coefficients, shifting attention from exact support recovery to estimation and prediction.

Computationally, we adapt the ``cycBAR'' algorithm of \citet{kawaguchi2021scalable} to the high-dimensional VAR setting. Its coordinatewise thresholding updates avoid repeated inversion of large Gram matrices and remain feasible when the number of regressors exceeds the sample size. Monte Carlo experiments across high-dimensional VAR designs show that BAR reduces shrinkage bias on large coefficients and performs strongly in estimation and prediction relative to Lasso, adaptive Lasso, and SCAD.

Finally, we illustrate the method in an application to U.S. inflation forecasting. The empirical exercise is consistent with the advantages suggested by the theory and simulations: BAR yields a stable sparse representation and delivers competitive out-of-sample forecasts relative to standard penalized VAR approaches, with particularly strong performance at the one-month horizon.

A few notational conventions are used throughout. Bold lowercase and uppercase letters denote vectors and matrices. For a matrix $\bM$, $\norm{\bM}_{\max}$ and $\norm{\bM}_{\mathrm{op}}$ denote the elementwise sup- and operator norms, and $\lambda_{\min}(\bM)$ and $\lambda_{\max}(\bM)$ its extreme eigenvalues. For a vector $\bv$, $\norm{\bv}_r$ is the usual $\ell_r$ norm and $\norm{\bv}_0$ its number of nonzero entries. For a scalar random variable $Z$, $\|Z\|_{L_q}:=(\mathbb E|Z|^q)^{1/q}$ denotes its $L_q$ norm. We write $a_T\lesssim b_T$ if there exists a constant $C>0$, independent of $(K,T)$, such that $a_T\le Cb_T$ uniformly in $(K,T)$ while $a_T\asymp b_T$ if both inequalities hold. All asymptotic statements are joint in $K,T\to\infty$ unless stated otherwise.

\section{Modeling Set up}\label{sec_model_setup}
Let $\by_1,\ldots,\by_T$ be a $K$-dimensional multiple time series process, where $\by_t=(y_{1,t},\ldots, y_{K,t})^{'}$. 
We assume $\by_t$ is generated by a stable VAR($p$) process as
\begin{equation} \label{eq:dgp}
\by_t=\bA_1 \by_{t-1}+\cdots+\bA_{p}\by_{t-{p}}+\bu_t, \qquad t=p+1,\ldots,T\;,
\end{equation}
where $\bA_1,\ldots,\bA_{p}$ are $K\times K$ transition matrices and $\bu_t$ is a $K\times 1$ vector of zero mean error terms with covariance $\bSigma_u=\mathbb{E}(\bu_t\bu_t')$.\footnote{Stability of the VAR implies: $\operatorname{det}\left(\bI_K-\sum_{\ell=1}^p \bA_{\ell} z^{\ell}\right)\neq 0, \quad \forall |z|\leq 1.$} We omit both the intercept and any polynomial time trend from the model to ease the notation, but the results we derive easily extend to those cases as well. The following assumptions are satisfied by the VAR in \eqref{eq:dgp}.

\begin{assumption}[Data generating process (DGP)]\label{ass_phidep}
Let $\bh_t:=(\by_t',\bu_t')'\in\mathbb R^{2K}$, $t\in\mathbb Z$, be covariance stationary. Suppose that there exist constants $q>4$, $C>0$, and $\alpha>0$, with $\alpha\neq 1/2-2/q$, such that:
\begin{enumerate}[label=(\roman*)]
    \item $\mathbb E(\bh_t)=0$; $\mathbb E(\bu_t\mid\by_{t-1},\ldots,\by_{t-p})=0$; $\max_{1\le j\le 2K}\mathbb E(|h_{j,t}|^q)\le C$.
    \item The process is causally generated as $\bh_t=G(\ldots,\be_{t-1},\be_t)=G(\mathcal F^t)$, where $\{\be_t\}_{t\in\mathbb Z}$ is an iid sequence in $\mathbb R^{2K}$ and $G:\mathcal E^\infty\to\mathbb R^{2K}$ is measurable.\footnote{Here $\mathcal F^t:=(\ldots,\be_{t-1},\be_t)$ is the realized
sequence of innovations observed up to time $t$. If $\mathcal E$
denotes the set of possible values of each innovation $\be_t$, then
$\mathcal E^\infty$ denotes the set of all possible semi-infinite
sequences with elements in $\mathcal E$. Thus,
$\mathcal F^t\in\mathcal E^\infty$, and
$G:\mathcal E^\infty\to\mathbb R^{2K}$ maps each possible innovation
history into the corresponding value of $\bh_t$.}
    \item Let $\mathcal F^{t,0}:=(\ldots,\be_{-1},\be_0^*,\be_1,\ldots,\be_t)$, where $\be_0^*$ is an independent copy of $\be_0$, and set $\bh_{t,0}:=G(\mathcal F^{t,0})$. For $j=1,\ldots,2K$, define $\delta_{q,j,t}:=\|h_{j,t}-h_{j,t,0}\|_{L_q}$, $\Delta_{q,j,m}:=\sum_{t=m}^\infty\delta_{q,j,t}$, and $\|\bh_{j,\bullet}\|_{q,\alpha} :=\sup_{m\ge0}(m+1)^\alpha\Delta_{q,j,m}$, the latter being the dependence adjusted $q$-norm (DANq) for the process $\bh_{j, \bullet}=\{h_{j,t}\}_{t\in \mathbb{Z}}$, with $\Psi_{q,\alpha}:=\max_j\|\bh_{j,\bullet}\|_{q,\alpha}$. To control the largest coupling effect across the system, also let $\omega_{q,t}:= \big\|\|\bh_t-\bh_{t,0}\|_\infty\big\|_{L_q}$, $\Omega_{q,m}:=\sum_{t=m}^\infty\omega_{q,t}$, and $\mathcal H_{q,\alpha,T}:=\sup_{m\ge0}(m+1)^\alpha\Omega_{q,m}$. Set $r:=q/2>2$ and let $\alpha_1:=1$ if $\alpha>1/2-1/r$, while $\alpha_1:=r/2-r\alpha$ if $0<\alpha<1/2-1/r$. Assume \begin{align}
\Psi_{q,0}\vee\Psi_{q,\alpha} &\le C, \label{eq:coordinate_dan_bound}\\ \mathcal H_{q,0,T}\mathcal H_{q,\alpha,T}\log(Kp+1) &= o\!\left(T^{1/2-\alpha_1/r}\right).
\label{eq:linfty_dan_growth} \end{align}
    \item The number of regressors in each equation satisfies $Kp=O(T^\nu)$ for some fixed $\nu>0$.
\end{enumerate}
\end{assumption}

\noindent Assumption \ref{ass_phidep}(i) imposes conditional mean-zero innovations and uniformly bounded $q$th moments. Parts (ii)--(iii) formulate short-range temporal dependence through the physical-dependence framework of \citet{wu2005nonlinear}, under which the stationary causal process is also ergodic. This framework is useful because it accommodates a broad range of nonlinear and conditionally heteroskedastic time-series dynamics while retaining a tractable innovation-based measure of dependence.\footnote{Examples and primitive sufficient conditions for familiar VARMA- and multivariate-GARCH-type processes are given in \citet[][Online Appendix C2]{chernozhukov2021lasso}.} The coupling mechanism replaces one innovation by an independent copy and measures how strongly this perturbation propagates through the future of the process. Boundedness of $\Psi_{q,\alpha}$ in \eqref{eq:coordinate_dan_bound} implies the remaining cumulative effect must decay at least polynomially, uniformly over the individual coordinates of the process. The quantity $\mathcal H_{q,\alpha,T}$, which is allowed to grow with the dimension, strengthens this coordinatewise control by tracking the largest coupling effect across the entire system. This additional control is needed for the joint high-dimensional concentration argument used later for the empirical process and sample Gram matrix; see Lemma \ref{lem_EPB}. Condition \eqref{eq:linfty_dan_growth} restricts the combined growth of the cumulative maximal coupling effect $\mathcal H_{q,0,T}$, its temporal persistence as measured by $\mathcal H_{q,\alpha,T}$, and the dimension through $\log(Kp+1)$, relative to the sample size. This sufficient condition is adapted from the vector-valued maximal inequality of \citet{zhang2017gaussian} and its second-order implementation in \citet{margaritella2025precision}. Part (iv) permits polynomial dimensional growth and does not require $Kp<T$; in particular, $Kp/T\to\infty$ is allowed when $\nu>1$. Because $p$ is fixed (see Remark \ref{rmk_laglength} below), all dimensional growth is driven by $K$. A fuller discussion, including a simple coupling example and the relation between coordinatewise and maximal dependence, is given in Appendix \ref{app:physical_dependence_details}.

\begin{remark}[VAR order $p$]\label{rmk_laglength}
We take $p$ to be fixed, known, and small. As usual, a finite-order VAR may be viewed as a working approximation to a more general VAR$(\infty)$ representation; our theoretical results are conditional on the chosen truncation order and do not address the approximation error from omitted higher lags. Keeping $p$ small is nevertheless natural in high-dimensional VARs because cross-sectional propagation already generates rich marginal dynamics: the final-equation representation implies that a VAR$(p)$ induces univariate ARMA orders up to $(Kp,(K-1)p)$, see \citet{cubadda2009} and the references therein. Thus, even $p=1$ can accommodate substantial persistence when $K$ is large. 
\end{remark}

\begin{assumption}[Row-wise Approximate Sparsity]\label{ass_sparsity}
 Each row of the concatenated coefficient matrix $\boldsymbol{\mathcal{A}}:=\left[\bA_1 \cdots \bA_p \right]\in \mathbb{R}^{K\times Kp}$ is approximately sparse, that is, for $i=1,\ldots,K$, \begin{align*}     \max_i\sum_{j=1}^{Kp} |\mathcal{A}_{ij}|^{q_s}=  \max_i \sum_{\ell=1}^p \sum_{j=1}^K |A_{\ell,ij}|^{q_s}\leq s_0, \quad q_s\in [0,1). \end{align*} \end{assumption}
 
Assumption \ref{ass_sparsity} allows for approximate, rather than exact, row-wise sparsity. When $q_s=0$, it reduces to exact sparsity, with $s_0$ bounding the number of nonzero coefficients in each row of $\boldsymbol{\mathcal A}$. For $q_s>0$, coefficients need not be exactly zero, and $s_0$ instead controls the overall weak-sparsity level of each row: for fixed $q_s$, larger values of $s_0$ allow more or larger non-negligible coefficients. The parameter $q_s$ determines how much of this signal may be spread over small coefficients, with smaller values corresponding to stronger sparsity restrictions. In the VAR setting, this row-wise formulation implies that each variable can be strongly influenced by only a limited number of the $Kp$ lagged regressors, leading to an approximately sparse interaction structure \citep[see also][]{krampe2021sparsity}.  

Because BAR is applied equation-by-equation, we shall without loss of generality focus on a single equation  of the system in \eqref{eq:dgp}, i.e., the $i$-th component of $\by_t$ as $y_{i,t}=\bx_t'\bbeta_i+u_{i,t},$ for $\bx_t=(\by_{t-1}',\ldots,\by_{t-p}')'$ and stack this along the time dimension as 
\begin{align} \label{stacked_dgp}
\by_i=\bX\bbeta_i+\bu_i,\quad i=1,\ldots,K,    
\end{align}
where $\by_i, \bu_i\in\mathbb{R}^{T-p}$, $\bX=(\bx_{p+1}',\ldots,\bx_T')'\in\mathbb{R}^{(T-p)\times Kp}$. Since each column of $\bX$ is one lag of one variable across time, we later write $\bX_j$ for the $j$th column of $\bX$, $j=1,\ldots,Kp.$ Since the first $p$ observations are used to construct the lagged regressors, the effective sample size is originally $T-p$. To simplify notation, from this point onward we relabel this effective sample size as $T$. Since $p$ is fixed (see Remark \ref{rmk_laglength}), this relabeling has no asymptotic consequence.

The transpose of the $i$th row of $\boldsymbol{\mathcal A}$ is precisely the coefficient vector $\bbeta_i$ in \eqref{stacked_dgp}; hence Assumption \ref{ass_sparsity} implies $\|\bbeta_i\|_{q_s}^{q_s}\le s_0$. Moreover, because $y_{i,t}$, $u_{i,t}$, and each component of $\bx_t$ are coordinates or lags of $\bh_t$, the single-equation representation inherits the physical-dependence conditions in Assumption \ref{ass_phidep}. Henceforth, we consider an arbitrary equation $i$ and suppress its index. The assumptions and subsequent bounds are stated with constants that do not depend on $i$, so the same equationwise results apply to every VAR equation.

\begin{figure}[htbp]
    \centering
    \includegraphics[width=0.6\textwidth,trim=0.1cm 0.1cm 0.15cm 0.45cm,clip]{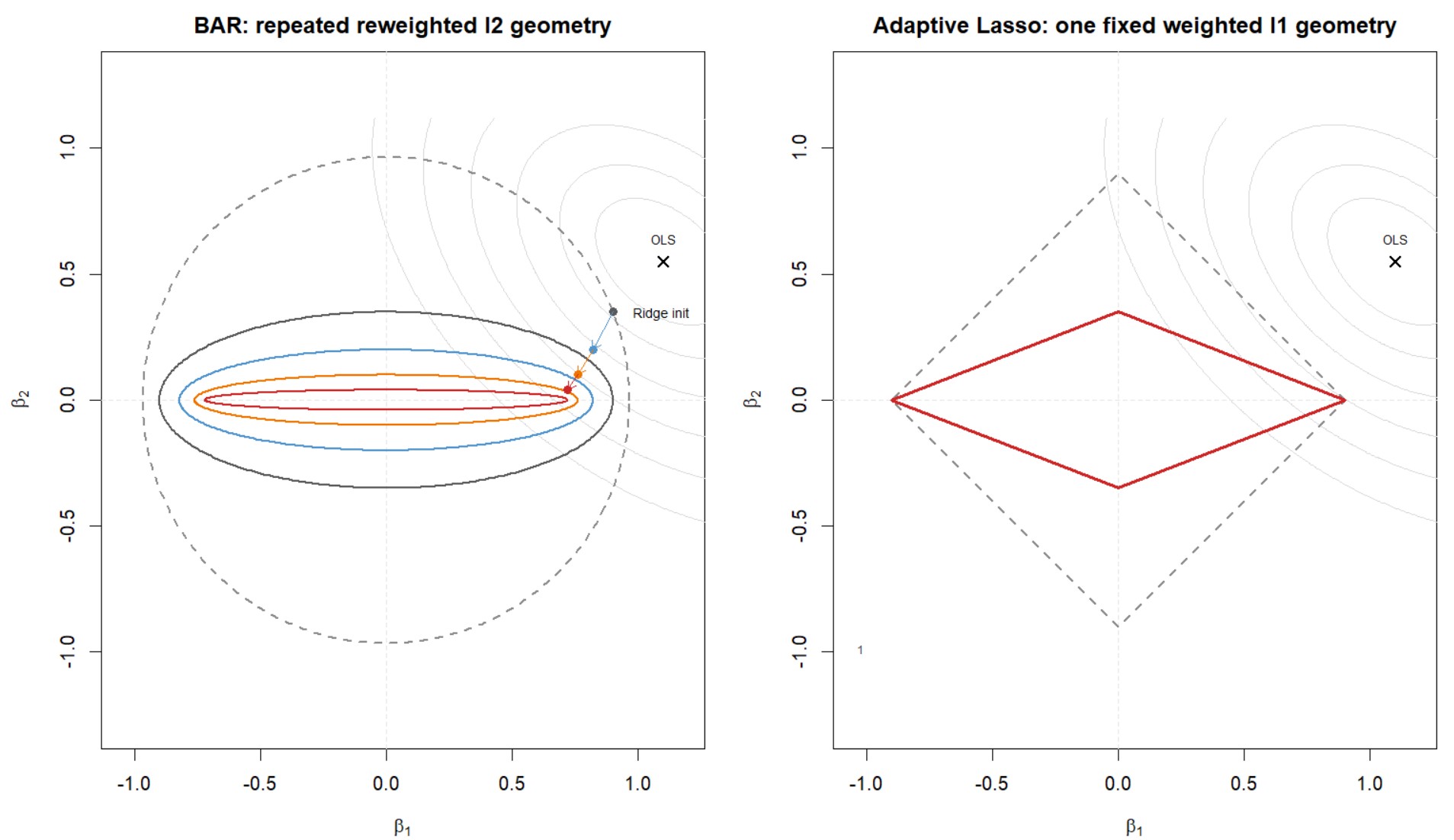}
    \caption{\footnotesize Stylized two-dimensional geometry of BAR and Adaptive Lasso. Left: starting from a Ridge initialization, BAR generates a sequence of reweighted $\ell_2$-type penalty contours. The ordinary Ridge contour is spherical, while successive BAR contours become increasingly compressed in the direction of the smaller coordinate. Right: Adaptive Lasso uses a single weighted $\ell_1$ penalty contour based on a initializing estimator. Relative to the ordinary Lasso diamond, the weighted diamond is more compressed in directions associated with larger penalty weights, but its geometry is fixed once the initializer weights are chosen.} 
    \label{fig:bar_adalasso_geometry}
\end{figure}

BAR builds a sequence of weighted Ridge regressions, starting from an ordinary Ridge estimator and using each iterate to determine the weights in the next step. For $\tau\ge1$, define
\begin{align}
\hat{\bbeta}^{(0)}
&=\argmin_{\bbeta}\left\{T^{-1}\|\by-\bX\bbeta\|_2^2+\xi\sum_{j=1}^{Kp}\beta_j^2\right\},\qquad
\hat{\bbeta}^{(\tau)}=g\{\hat{\bbeta}^{(\tau-1)}\}, \label{eq_recursion}\\
g(\widetilde{\bbeta})
&=\argmin_{\bbeta}\left\{T^{-1}\|\by-\bX\bbeta\|_2^2+\lambda\sum_{j=1}^{Kp}\frac{\beta_j^2}{\widetilde\beta_j^2}\right\},\qquad
\hat{\bbeta}^{BAR}=\lim_{\tau\to\infty}\hat{\bbeta}^{(\tau)}. \label{eq_reweight}
\end{align}
Equations \eqref{eq_recursion}--\eqref{eq_reweight} give the original
iteratively reweighted Ridge formulation of BAR proposed by
\citet{dai2018broken}. 
Here $\xi$ and $\lambda$ are tuning parameters, and the closed-form Ridge expressions follow directly from the first-order conditions. The reweighting scheme, similar in spirit to the adaptive Lasso of \citet{zou2006adaptive}, assigns increasingly large weights to small coefficients while leaving sufficiently large coefficients comparatively unshrunk. This breaks away from the ordinary Ridge path and induces sparsity in the limit; see Figure \ref{fig:bar_adalasso_geometry} for an illustration of the geometry.

Generically, neither the initial Ridge estimator $\hat{\bbeta}^{(0)}$ nor any finite iterate $\hat{\bbeta}^{(\tau)}$ in \eqref{eq_recursion} contains exact zeros, since the weighted $\ell_2$ problems do not induce sparsity. Exact zeros therefore arise only at the limiting fixed point of the BAR recursion. Consequently, a numerical implementation of the original recursion requires a cutoff to classify sufficiently small finite-iterate coefficients as zero.

Direct implementation of \eqref{eq_reweight} can though also be numerically unstable because the weights become very large when some elements of the previous iterate approach zero. While a small perturbation of the weights can alleviate this problem, \citet{dai2018broken} proposed a more numerically stable reformulation based on rescaling the regressors by the current coefficient estimates, thereby avoiding the explicit formation of unstable weights.\footnote{$ g(\widetilde{\bbeta}) = \bGamma(\widetilde{\bbeta}) \left( T^{-1}\widetilde{\bX}(\widetilde{\bbeta})' \widetilde{\bX}(\widetilde{\bbeta}) +\lambda\bI \right)^{-1} T^{-1}\widetilde{\bX}(\widetilde{\bbeta})'\by$, where $\bGamma(\widetilde{\bbeta}) = \operatorname{diag}(\widetilde{\beta}_1,\ldots, \widetilde{\beta}_{Kp})$ and $\widetilde{\bX}(\widetilde{\bbeta}) = \bX\bGamma(\widetilde{\bbeta})$; see \citet{dai2018broken}.}
However, a direct implementation of this formulation entails
the repeated inversion of $(Kp)\times(Kp)$ matrices and can therefore become computationally burdensome in high dimensions.
This was not a central issue in \citet{dai2018broken}, where the number of regressors was assumed to be smaller than the sample size or was reduced through preliminary screening. In our setting, Assumption \ref{ass_phidep}(iv) allows $Kp$ to be much larger than $T$. A direct implementation would then require repeated inversion of dense $(Kp)\times(Kp)$ matrices, with a computational cost of order $O((Kp)^3)$ per iteration under standard matrix-inversion routines.\footnote{For example, Gaussian elimination or LU decomposition for a dense $n\times n$ matrix requires $2n^3/3+O(n^2)$ operations. Here, $n=Kp$.}

To avoid these inversions, we adopt the coordinatewise fixed-point strategy underlying ``cycBAR'', as designed in \cite{kawaguchi2021scalable}. Rather than solving the complete weighted Ridge problem in \eqref{eq_reweight} at every iteration, we update each coordinate directly through the corresponding BAR fixed-point equation. For the least-squares loss considered here, this equation admits an explicit thresholding representation with a closed-form outer-root solution. The resulting procedure produces exact zeros at each iteration and avoids the inversion of large Gram matrices. Its pseudo-code is summarized in Algorithm~\ref{alg:BAR_CD}.
{\footnotesize
\begin{algorithm}
\caption{Pseudo-code of BAR Coordinatewise Fixed-Point Thresholding}
\label{alg:BAR_CD}  
\hrule
\begin{algorithmic}[1]
    \State Inputs: $\bm{y}, \bm{X}$ (Data); $\xi, \lambda$ (Tuning Parameters); $N_{max}$ (Max nr iter); $\epsilon$ (Tolerance);
    \State Initialize \(\bm{\hat{\beta}}^{(0)} = (T^{-1}\bm{X' X} + \xi \bm{I})^{-1} (T^{-1}\bm{X' y})\); \hfill (Ridge estimator)
    \For{$n = 1$ to $N_{\max}$} \hfill (Fixed-point iteration)
        \For{$j = 1$ to $Kp$} \hfill (Coordinatewise fixed-point update)
            \State compute the partial residual: $\br_j^{[n]}=\by-\bX\hat{\bbeta}^{(n-1)}+\bX_j\hat\beta_j^{(n-1)}
            =\by-\bX_{-j}\hat{\bbeta}_{-j}^{(n-1)}$;
            \State compute the normalized score: $b_j^{[n]}= T^{-1}\bm{X}_j' \bm{r_j}^{[n]}$;
            \State compute the activation threshold: $\mathcal T_j(\lambda) = 2\sqrt{\lambda (T^{-1}\bm{X}_j' \bm{X}_j)}$;
            \If{$|b_j^{[n]}| \geq \mathcal T_j(\lambda)$} \hfill (Thresholding update)
                \State $\hat{\beta}_j^{(n)} = \dfrac{b_j^{[n]} + \text{sign}\left(b_j^{[n]}\right)\sqrt{\left(b_j^{[n]}\right)^2 - 4\lambda (T^{-1}\bm{X}_j' \bm{X}_j)}}{2T^{-1}\bm{X}_j' \bm{X}_j}$;
            \Else
                \State $\hat{\beta}_j^{(n)} = 0$;
            \EndIf
        \EndFor
        \If{$\|\bm{\hat{\beta}}^{(n)} - \bm{\hat{\beta}}^{(n-1)}\| \leq \epsilon$} \hfill (Convergence check)
            \State Break \hfill (Stop upon convergence)
        \EndIf
    \EndFor
    \State Upon convergence, set
$\bm{\hat{\beta}}_{BAR}=\bm{\hat{\beta}}^{(n)}$; \hfill (BAR estimator)
\end{algorithmic}
\hrule
\end{algorithm}
}

Few more details on Algorithm \ref{alg:BAR_CD} are in order. The key idea is to start from the weighted-Ridge recursion, assume it converges, and characterize what any nonzero limiting coefficient must satisfy. Indeed, suppose that the recursion in \eqref{eq_recursion}--\eqref{eq_reweight} converges to $\hat{\bbeta}_{BAR}$ and let $j\in\hat S:=\{j:\hat\beta_{BAR,j}\neq0\}$. Define the partial residual $\br_j(\hat{\bbeta}_{BAR}) :=\by-\bX\hat{\bbeta}_{BAR}+\bX_j\hat\beta_{BAR,j}$, the corresponding normalized score $b_j(\hat{\bbeta}_{BAR}) :=T^{-1}\bX_j'\br_j(\hat{\bbeta}_{BAR})$, and $a_j:=T^{-1}\bX_j'\bX_j$. At iteration $\tau$ of the original weighted-Ridge recursion, the weights are fixed at their values from iteration $\tau-1$, and $\hat{\bbeta}^{(\tau)}$ solves the corresponding weighted-Ridge problem. Hence, for coordinate $j$, its first-order condition is $-T^{-1}\bX_j'(\by-\bX\hat{\bbeta}^{(\tau)}) +\lambda\hat\beta_j^{(\tau)}/\{\hat\beta_j^{(\tau-1)}\}^2=0$.\footnote{Here $\tau$ indexes the original reweighting recursion in \eqref{eq_recursion}--\eqref{eq_reweight}, not the cyclic coordinate updates in Algorithm \ref{alg:BAR_CD}.} If
$\hat{\bbeta}^{(\tau)}\to\hat{\bbeta}_{BAR}$ as $\tau\to\infty$ and
$\hat\beta_{BAR,j}\neq0$, then also
$\hat\beta_j^{(\tau-1)}\to\hat\beta_{BAR,j}$, so that $\lambda
\frac{\hat\beta_j^{(\tau)}}
{\{\hat\beta_j^{(\tau-1)}\}^2}
\longrightarrow
\frac{\lambda}{\hat\beta_{BAR,j}}.$ Thus, passing to the limit $\tau\to\infty$ gives the active-coordinate fixed-point identity
\begin{align}
\label{eq:fixed_point_identity}
T^{-1}\bX_j'(\by-\bX\hat{\bbeta}_{BAR}) = \lambda/\hat\beta_{BAR,j}. \end{align} Combining this with $T^{-1}\bX_j'(\by-\bX\hat{\bbeta}_{BAR}) =b_j(\hat{\bbeta}_{BAR})-a_j\hat\beta_{BAR,j}$ yields a quadratic equation $a_j\hat\beta_{BAR,j}^2-b_j(\hat{\bbeta}_{BAR})\hat\beta_{BAR,j}+\lambda=0$ which is solved as \begin{align} \hat\beta_{BAR,j} = \frac{ b_j(\hat{\bbeta}_{BAR}) + \operatorname{sign}\!\{b_j(\hat{\bbeta}_{BAR})\} \sqrt{b_j(\hat{\bbeta}_{BAR})^2-4\lambda a_j} }{ 2a_j }. \label{eq:coordinate_fixed_quadratic} \end{align}
A real nonzero fixed point exists only if
$|b_j(\hat{\bbeta}_{BAR})|\ge2\sqrt{\lambda a_j}$, which is precisely the coordinate activation threshold $\mathcal T_j(\lambda)$ in Algorithm \ref{alg:BAR_CD}. When $|b_j(\hat{\bbeta}_{BAR})|>2\sqrt{\lambda a_j}$, the quadratic has two roots. Following \citet{kawaguchi2021scalable}, we select the outer root; its stability under the scalar fixed-point iteration is shown in Appendix \ref{app:coordinate_update_derivation}. Algorithm \ref{alg:BAR_CD} applies this update coordinate by coordinate using the current normalized score; if the threshold is not met, the coefficient is set to zero. Another useful feature of the cyclic update is that it produces exact zeros directly through the activation rule. This is unlike the original weighted-Ridge BAR recursion, whose finite iterates remain dense and therefore require an arbitrary numerical cutoff to induce sparsity. Hence, the algorithm searches directly for a solution of the BAR fixed-point equations without solving a full weighted-Ridge problem at every iteration.

Further implementation details, including reduced-space computation of the Ridge initializer and the roles of $N_{\max}$ and $\epsilon$, are reported in Appendix \ref{app:implementation_details}.

\begin{remark}[Standardization]\label{rmk_stand}
In the theoretical analysis below, the predictors are understood to be population centered and scaled, so that $\mathbb E(x_{j,t}^2)=1$. Under Assumption \ref{ass_phidep} and the later Lemma \ref{lem_EPB}, $\max_j|T^{-1}\bX_j'\bX_j-1|=o_p(1)$, so replacing the sample second moments by one does not affect the asymptotic rates. In implementation, the predictors are sample centered and scaled so that $T^{-1}\bX_j'\bX_j=1$, or equivalently $\|\bX_j\|_2^2=T$, exactly. Then the activation threshold in Algorithm \ref{alg:BAR_CD} becomes $\mathcal T_j(\lambda)=2\sqrt{\lambda}$ for all $j$, and the active-coordinate update reduces to $\{b_j^{[n]}+\operatorname{sign}(b_j^{[n]}) [(b_j^{[n]})^2-4\lambda]^{1/2}\}/2$. In particular, every active coordinate satisfies $|\hat\beta_j^{(n)}|\ge\sqrt{\lambda}$. Algorithm \ref{alg:BAR_CD} is written in its unnormalized form only to make the general mechanics transparent. \end{remark} Henceforth, the predictors are understood to be population standardized.

\begin{remark}[Grouping effect]\label{rmk_grouping}
The BAR fixed-point identity also yields a finite-sample grouping property. Let $\hat\rho_{jk}:=T^{-1}\bX_j'\bX_k$ denote the sample correlation between the standardized regressors $\bX_j$ and $\bX_k$. As shown in Proposition \ref{prop:grouping} of Appendix \ref{app:grouping_property}, for any $j,k\in \hat S:=\{j:\hat\beta_j\neq0\}$, \begin{align}
\left| \frac{1}{\hat\beta_{BAR,j}} - \frac{1}{\hat\beta_{BAR,k}} \right| \le \frac{\|\by\|_2}{\lambda\sqrt T} \sqrt{2(1-\hat\rho_{jk})}. \label{eq:grouping_remark} \end{align} Thus, the inverse estimated coefficients become close as $\hat\rho_{jk}$ approaches one. An analogous result holds as $\hat\rho_{jk}$ approaches minus one, in which case the inverse coefficients become close after a sign reversal. This property is particularly relevant for persistent VARs, where different lags of the same variable are often strongly correlated. \end{remark}

\par Algorithm~\ref{alg:BAR_CD} described the coordinatewise fixed-point
thresholding implementation of BAR for fixed tuning parameters
$\xi$ and $\lambda$.
In practice, the BAR estimator -along with its initial Ridge warm start- is applied over a grid of penalty parameters, as described in Algorithm \ref{alg:BAR_path}.

\begin{algorithm}
\caption{BAR Tuning Path over a Grid of Penalties}
\label{alg:BAR_path}
\hrule
\begin{algorithmic}[1]
    \State Inputs: $\by,\bX$, Ridge parameter $\xi$, $N_{\max}$, tolerance $\epsilon$, and a grid $\Lambda=\{\lambda^{(1)},\ldots,\lambda^{(G)}\}$
    \For{$g=1,\ldots,G$}
        \State Run Algorithm \ref{alg:BAR_CD} at $\lambda^{(g)}$ and obtain $\hat{\bbeta}_{BAR}(\lambda^{(g)})$
        \State Set $\hat S(\lambda^{(g)})=\{j:\hat\beta_{BAR,j}(\lambda^{(g)})\neq0\}$ and $RSS(\lambda^{(g)})=\|\by-\bX\hat{\bbeta}_{BAR}(\lambda^{(g)})\|_2^2$
        \State Evaluate an information criterion $\mathcal C(\lambda^{(g)})$
    \EndFor
    \State Select $\hat\lambda=\argmin_{\lambda\in\Lambda}\mathcal C(\lambda)$ and return $\hat{\bbeta}_{BAR}(\hat\lambda)$ and $\hat S(\hat\lambda).$
\end{algorithmic}
\hrule
\end{algorithm}
The computational grid in Algorithm \ref{alg:BAR_path} ranges from a lower endpoint anchored to the theoretical rate to a data-driven upper endpoint, specifically: \begin{align}\label{eq:bar_grid_endpoints} \lambda^{(1)}=10^{-3}\frac{\log(Kp)}{T},\qquad \lambda^{(G)}=\frac14\max_{1\le j\le Kp}\frac{(b_j^{[1]})^2}{T^{-1}\bX_j'\bX_j}. \end{align} The lower endpoint follows from the empirical-process rate derived below in Lemma \ref{lem_EPB}: because the BAR threshold is of order $\sqrt{\lambda}$, the smallest statistically meaningful penalty is of order $\log(Kp)/T$. Here, we use a deliberately small multiplier ($10^{-3}$) so that the computational path also includes relatively dense models. The upper endpoint is the first-iteration activation boundary; penalties strictly above $\lambda^{(G)}$ yield no active coordinate at the first iteration. Under standardization, it reduces to $\lambda^{(G)}=\frac14\max_j(b_j^{[1]})^2$. The Ridge parameter $\xi$ determines the dense initializer and therefore influences the first-iteration normalized scores, whereas $\lambda$ directly controls thresholding and sparsification. 

The tuning parameters $\xi$, and in particular $\lambda$, are central to both the computation and the theory of BAR. Algorithm \ref{alg:BAR_path} chooses them by minimizing an information criterion $\mathcal{C}(\lambda)$ along the corresponding tuning path. In Section \ref{sec_lambda_ic}, and supported by the simulation evidence reported in Section \ref{sec:simulations}, we advocate for the use of the Bayesian Information Criterion (BIC) for the Ridge initializer parameter $\xi$, and the Extended BIC \citep[EBIC;][]{ebic_chen2008} for the sparsity-inducing penalty $\lambda$. This combination works well in simulations: BIC yields a numerically stable initializing estimator, whereas EBIC tends to avoid the extreme ends of the BAR path and more often selects penalties in the empirically relevant interior region. More on this in Section \ref{sec_lambda_ic}.

%%%%%%%%%%%%%%%% 

\section{BAR High-Dimensional Theory \& Error Bounds}
Given a BAR estimate $\hat{\bbeta}_{BAR}\in\mathbb R^{Kp}$, we assess its quality through prediction and estimation consistency. Prediction consistency $\|\bX(\hat{\bbeta}_{BAR}-\bbeta)\|_2^2$ concerns how well the fitted model reproduces the systematic component of the data. Estimation consistency, instead, concerns how close $\hat{\bbeta}_{BAR}$ is to the true parameter vector $\bbeta$ as measured by $\|\hat{\bbeta}_{BAR}-\bbeta\|_r$, $r\in\{1,2\}$. In high-dimensional settings, the two need not coincide, so both criteria are considered.

\par Under approximate sparsity (Assumption \ref{ass_sparsity}), exact support recovery is not a natural main objective, since $\bbeta$ may contain many small but nonzero coefficients. For a fixed tuning parameter $\lambda\geq0$, we therefore write $\hat S:=\hat S(\lambda)$ for the support selected by BAR and define the corresponding population ``effective support'' as
\begin{align}\label{eq_support}
    S_{\lambda}:=\left\{j:|\beta_j|>2\sqrt{\lambda}\right\}.
\end{align}
We denote its cardinality by $|S_{\lambda}|$ and its complement by
$S_{\lambda}^c=\{1,\ldots,Kp\}\backslash S_{\lambda}$. The set
$S_{\lambda}$ collects the true coefficients lying above the BAR
activation threshold $\mathcal{T}_j(\lambda)$. For $\lambda=0$, it reduces to the ``exact support'' of $\bbeta$.\footnote{Note how approximate sparsity is fully compatible with the $\ell_0$-type thresholding geometry of BAR: the estimator need not assume that the DGP is exactly sparse, but only that coefficients below the threshold have sufficiently small aggregate mass.} The next lemma formalizes the two roles played by $S_{\lambda}$: controlling the number of true coefficients above the BAR threshold scale and, at the same time, the aggregate mass of those below it. Both quantities will indeed enter the BAR error bounds and therefore control for them is needed.
\begin{lemma}[Effective support and tail bounds]\label{lem_tailsize}
Given Assumption \ref{ass_sparsity}, for every $q_s\in[0,1)$ and
$\lambda>0$,
\begin{align}
|S_{\lambda}| = \sum_{j=1}^{Kp}\mathbb{1}\{|\beta_j|>2\sqrt{\lambda}\} \;\lesssim\; s_0\,\lambda^{-q_s/2}, \qquad \|\bbeta_{S_{\lambda}^c}\|_r \;\lesssim\; s_0^{1/r}\,\lambda^{\frac{1}{2}-\frac{q_s}{2r}}, \quad r\in\{1,2\}. \label{eq:weak_sparsity_unified}
\end{align}
\end{lemma}
The bounds in Lemma \ref{lem_tailsize} clarify the role of the weak-sparsity parameters from Assumption \ref{ass_sparsity}. The overall amount of signal is controlled by $s_0$, while $q_s$ determines how much of it may be spread over small coefficients. Accordingly, the bounds allow both the effective support and the aggregate mass below the threshold to increase with $s_0$ and $q_s$. The tuning parameter $\lambda$ determines the trade-off between the two: a smaller $\lambda$ lowers the cutoff $2\sqrt{\lambda}$ and therefore increases $|S_\lambda|$, but at the same time reduces $\|\bbeta_{S_\lambda^c}\|_r$, since fewer coefficients remain below the cutoff. For $q_s=0$ the weak-sparsity condition reduces to exact sparsity, $\|\bbeta\|_0\le s_0$, and the bounds of Lemma \ref{lem_tailsize} continue to hold with $|S_\lambda|\le s_0$ and $\|\bbeta_{S_\lambda^c}\|_r\lesssim s_0^{1/r}\sqrt{\lambda}$. The proof is given in the \autoref{app:proofs}.

To make these bounds operational, however, we need to determine the appropriate size of $\lambda$. The cutoff $2\sqrt{\lambda}$ defining $S_\lambda$ in population is also the BAR activation threshold $\mathcal T_j(\lambda)$ in sample. Since BAR activates coordinate $j$ according to the normalized score $b_j^{[n]}=T^{-1}\bX_j'\br_j^{[n]}$, $\mathcal T_j(\lambda)$ should be large enough that a coordinate is not activated because of stochastic terms contained in $b_j^{[n]}$
pushing its absolute value erroneously over $\mathcal T_j(\lambda)$. The next subsection decomposes $b_j^{[n]}$ into the true signal and two additional random components generated by the empirical noise and the estimation errors in the remaining coordinates, see \eqref{eq_empdec}. We first derive the stochastic order of the empirical-noise component, which determines the appropriate order of $\lambda$, and then turn to the cross-design component in the following subsection.

%%%%%%%%%%%%
\subsection{Empirical Process and Gram-Matrix Concentration}\label{subsec_EPB}

The normalized score $b_j^{[n]}=T^{-1}\bX_j'\br_j^{[n]}$ is the main character in Algorithm \ref{alg:BAR_CD} (line 6) as its absolute value  decides whether the $j$th coefficient is activated or set to zero. It also reveals the two main stochastic ingredients required for the BAR analysis. Using $\br_j^{[n]}=\by-\bX_{-j}\hat{\bbeta}_{-j}^{(n-1)}$, substituting $\by=\bX\bbeta+\bu$, and via standardization: \begin{align} b_j^{[n]} = \beta_j - T^{-1}\bX_j'\bX_{-j} \bigl(\hat{\bbeta}_{-j}^{(n-1)}-\bbeta_{-j}\bigr) + T^{-1}\bX_j'\bu. \label{eq_empdec} \end{align} 
The three terms on the right hand side in \eqref{eq_empdec} are, respectively, the true $j$th signal $\beta_j$, a cross-design error term, and the empirical process (the ``noise''). For $j\in S_\lambda^c$, by definition $|\beta_j|\le\mathcal T(\lambda)$, so the true signal alone clearly does not exceed the BAR activation threshold.\footnote{We drop the $j$ subscript from $\mathcal T_j(\lambda)$ here as, under standardization, the threshold is the same for all $j$.} Thence, a coordinate outside the effective support can cross the activation threshold only because of the other two terms in \eqref{eq_empdec}. The second term, which at fixed point becomes $-T^{-1}\bX_j'\bX_{-j} (\hat{\bbeta}_{BAR,-j}-\bbeta_{-j})$,\footnote{This decomposition is written at a generic iteration to reflect the actual BAR update. At a fixed point, $\hat{\bbeta}^{(n-1)}=\hat{\bbeta}_{BAR}$, so the cross-design term becomes $-T^{-1}\bX_j'\bX_{-j}(\hat{\bbeta}_{BAR,-j}-\bbeta_{-j})$.} reflects how estimation errors in the remaining coordinates can feed into the $j$th normalized score through the off-diagonal entries of the sample Gram matrix $\hat{\bSigma}_x:=T^{-1}\bX'\bX$. The last term, $T^{-1}\bX_j'\bu$, is instead the direct contribution of the noise.  To determine the stochastic scale at which $\mathcal T(\lambda)$ should be set, we consider first the last term. Uniformly over the coordinates, the activation threshold should be at least of the same order as $\|T^{-1}\bX'\bu\|_\infty$; otherwise coordinates may be activated simply because the empirical noise term is large, rather than because of its true signal. Since $\mathcal T(\lambda)=2\sqrt{\lambda}$, this suggests choosing $\lambda$ on the order of $\|T^{-1}\bX'\bu\|_\infty^2$.

The cross-design error term requires a different type of control. It depends on the geometry of the sample Gram matrix and on the estimation error itself, and therefore cannot simply be absorbed into the penalty calibration as for the noise. Let $\bSigma_x:=\mathbb E(\bx_t\bx_t')$ denote the population counterpart of $\hat{\bSigma}_x$. The strategy pursued below is to first localize the estimation error to a sufficiently small support and then impose restricted curvature only along the relevant directions. This requires control of the sample Gram matrix in addition to control of the empirical process. Thus, the two stochastic objects needed for the subsequent analysis are $T^{-1}\bX'\bu$ and $\hat{\bSigma}_x-\bSigma_x$. Under Assumption \ref{ass_phidep}, both objects can be controlled jointly at the same high-dimensional rate, as done in the following lemma.

\begin{lemma}[Joint Empirical Process and Gram-matrix Concentration]
\label{lem_joint_concentration}
\label{lem_EPB}
\label{lem_verifyRE}
Under Assumption \ref{ass_phidep}, there exists a sufficiently large fixed constant $C_\chi>0$ such that the event \begin{align} \mathcal E_\chi := \left\{ \left\|T^{-1}\bX'\bu\right\|_\infty \vee \left\|\hat{\bSigma}_x-\bSigma_x\right\|_{\max} \le C_\chi\sqrt{\frac{\log(Kp)}{T}} \right\}, \label{eq:event_chi} \end{align} satisfies $\mathbb P(\mathcal E_\chi)\to1$ as $K,T\to\infty$. Then, $\left\|T^{-1}\bX'\bu\right\|_\infty \vee \left\|\hat{\bSigma}_x-\bSigma_x\right\|_{\max} = \mathcal O_p\left( \sqrt{\frac{\log(Kp)}{T}} \right)$.  \end{lemma}
The two bounds in Lemma \ref{lem_joint_concentration} play different roles. The bound on $\left\|T^{-1}\bX'\bu\right\|_\infty$ determines the stochastic scale of the BAR activation threshold, motivating $\lambda=C_\lambda\log(Kp)/T$, where $C_\lambda>0$ is a sufficiently large fixed constant.\footnote{In particular, $C_\lambda$ is chosen large enough relative to the concentration constant $C_\chi$ so that the BAR activation threshold dominates the empirical process contribution on $\mathcal E_\chi$. A stronger lower bound on $C_\lambda$ will be imposed later when controlling the number of selected coefficients.} The Gram-matrix deviation bound, instead, is useful for controlling the cross-design term in \eqref{eq_empdec}, since this term depends on how estimation errors in the remaining coordinates are transmitted through the off-diagonal elements of $\hat{\bSigma}_x$. The proof of Lemma \ref{lem_EPB} follows the second-order process construction of \citet{margaritella2025precision}, based on the vector-valued maximal inequality of \citet{zhang2017gaussian}.\footnote{Quadratic products inherit physical dependence at moment order $q/2$, so that the primitive condition $q>4$ provides the required moment order above two, while \eqref{eq:linfty_dan_growth} controls the cross-sectional maximum.}

Now, by Lemma \ref{lem_joint_concentration}, setting $\lambda=C_\lambda\log(Kp)/T$ within Lemma \ref{lem_tailsize} gives the following rates.  \begin{corollary}\label{cor_rates} If $\lambda=C_\lambda\log(Kp)/T$, then, for $q_s\in[0,1)$ and $r\in\{1,2\}$, \begin{align} |S_\lambda| &\lesssim s_0\left(\frac{T}{\log(Kp)}\right)^{q_s/2}, & \|\bbeta_{S_\lambda^c}\|_r &\lesssim s_0^{1/r} \left(\frac{\log(Kp)}{T}\right)^{\frac12-\frac{q_s}{2r}}.
\label{eq_smallcontr}
\end{align}
\end{corollary}
Corollary \ref{cor_rates} makes the dimensional scaling of Lemma \ref{lem_tailsize} explicit. Under Assumption \ref{ass_phidep}, $Kp=O(T^\nu)$ for some fixed $\nu>0$, so $\log(Kp)=O(\log T)$ and polynomial growth in the number of regressors enters these rates only through a logarithmic factor.\footnote{In particular, Assumption \ref{ass_phidep} allows $Kp$ to be much larger than $T$, as long as its growth is polynomial. Since $\lambda\asymp\log(Kp)/T$, the BAR activation threshold still vanishes as $T\to\infty$. Under approximate sparsity, this allows the number of coefficients above the threshold to increase with $T$, while the aggregate $\ell_1$ and $\ell_2$ mass of those remaining below the threshold can simultaneously vanish.} As $T/\log(Kp)$ increases, the effective support may grow at rate $s_0\{T/\log(Kp)\}^{q_s/2}$, while the $\ell_1$ and $\ell_2$ mass below the threshold shrink. Thus, larger samples allow more coefficients to enter the effective support, whereas faster growth in $Kp$ works in the opposite direction. These rates determine the sparsity growth allowed in the arguments that follow.

%%%% RESTRICTED EIGENV
\subsection{Restricted Curvature on the Support Union}
\begin{figure}[htbp]
    \centering
\includegraphics[width=0.6\textwidth, trim={2cm 3.5cm 2cm 6.5cm}, clip]{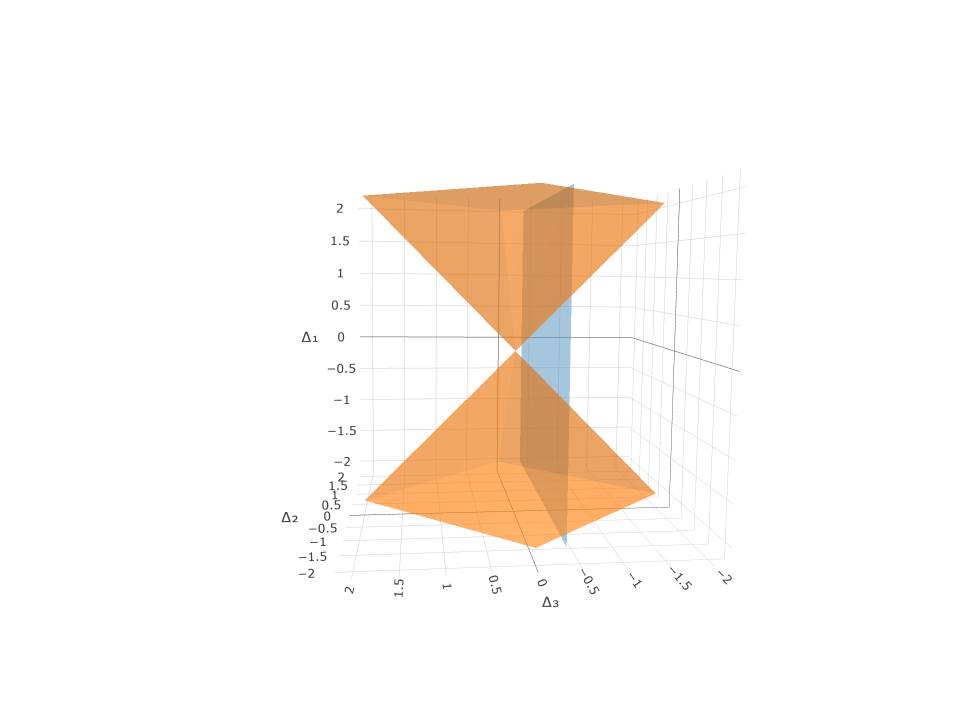}
\caption{\footnotesize Illustration of BAR error support restriction against the Lasso cone in $\mathbb{R}^3$. The blue plane represents the BAR support-union for $U=\{1,2\}$: since $\bDelta_{U^c}=-\bbeta_{U^c}$, the third coordinate is fixed at $\Delta_3=-\beta_3=-0.35$, while $\Delta_1$ and $\Delta_2$ remain free. The orange double cone represents the usual Lasso cone restriction, here illustrated by$|\Delta_2|+|\Delta_3|\le c^*|\Delta_1|$ with $c^*=1$. BAR localizes the error to a lower-dimensional affine set once $U$ is fixed, whereas the Lasso restricts the error to a cone of admissible directions.}
\label{fig:shell-cone}
\end{figure}

We now turn to the cross-design term in \eqref{eq_empdec}. At the BAR fixed point, let $\bDelta:=\hat\bbeta_{BAR}-\bbeta$. The term becomes $-T^{-1}\bX_j'\bX_{-j}\bDelta_{-j}$ and therefore cannot be controlled directly without
first exploiting some structure on the estimation error $\bDelta$. Unlike the Lasso, BAR does not restrict the estimation error to an $\ell_1$ cone. Its fixed-point rule instead has a hard-thresholding-type structure: coordinates below the activation threshold are set to zero, while active coordinates are mapped to a nonzero root. We exploit this structure to localize the estimation error to an affine set determined by the support union $U$, which we now define (see also Figure \ref{fig:shell-cone} for an illustration). 

Let $U:=S_{\lambda}\cup \hat S$. Recall $S_{\lambda}:=\{j:|\beta_j|>2\sqrt{\lambda}\}$
is the effective support under weak sparsity and
$\hat S:=\{j:\hat\beta_{BAR,j}\neq 0\}$
is the support selected by BAR. The set $U$ then contains both the coordinates that are truly large enough to matter and the coordinates that BAR actually selects. 
This construction is useful because outside such union, i.e., for every $j\in U^c$, the true coefficient lies below the threshold scale, since $j\notin S_{\lambda}$, while the estimated coefficient is exactly zero, as $j\notin \hat S$. Hence, the BAR error is exactly $\Delta_j=-\beta_j$ on $U^c$, making all the remaining uncertainty in the estimation error confined to the random set $U$.\footnote{A similar support-union argument appears in the analysis of iterative hard-thresholding methods, where the error is supported on the union of the target support and the support of the current iterate \citep{blumensath2009iterative}. In those methods, however, the size of the iterate (i.e., the maximum allowed of active coordinates) is fixed by construction.} The next lemma formalizes this discussion over the support characterization of the BAR error.
\begin{lemma}[BAR Error Representation on $U$]
\label{lem_BARshell}
Let $U:=S_\lambda\cup\hat S$, where $S_\lambda=\{j:|\beta_j|>2\sqrt{\lambda}\}$ and $\hat S=\{j:\hat\beta_{BAR,j}\neq0\}$. Then $\bDelta_{U^c}=-\bbeta_{U^c};\; \bDelta=\bDelta_U-\bbeta_{U^c}$. Since $U^c\subseteq S_\lambda^c$, for $r\in\{1,2\}$, $\|\bDelta_{U^c}\|_r \le \|\bbeta_{S_\lambda^c}\|_r;\; \|\bDelta\|_r \le \|\bDelta_U\|_r+\|\bbeta_{S_\lambda^c}\|_r$. In particular, plugging Corollary \ref{cor_rates} $\ell_r$-rates: \begin{align} \|\bDelta\|_r \le \|\bDelta_U\|_r + \mathcal O\!\left[ s_0^{1/r} \left(\frac{\log(Kp)}{T}\right)^{\frac{1}{2}-\frac{q_s}{2r}} \right], \qquad r\in\{1,2\}. \label{eq:shell_full_rates}
\end{align}
\end{lemma}
Lemma \ref{lem_BARshell} shows that outside $U$ the BAR estimation
error is completely determined by the true coefficients that lie below
the threshold, whose $\ell_r$ norms are already controlled
by Corollary \ref{cor_rates}. The remaining unknown part of the error is
therefore $\bDelta_U$, so the subsequent analysis can focus on
controlling the exact cardinality of $U$ and the size of $\bDelta_U$.

Controlling $\bDelta_U$ requires sufficient curvature of the quadratic loss component of the BAR objective along the coordinates in $U$. A lower bound on the smallest eigenvalue of $ \hat{\bSigma}_{x,UU}$ would ensure that small prediction errors translate into small estimation errors, as one would wish. This is the role of the restricted eigenvalue (RE) condition, typically employed in the literature. In a high-dimensional setting such curvature cannot be required globally, since $\hat{\bSigma}_x$ is singular whenever $Kp>T$, and any submatrix with more than $T$ columns is necessarily singular too. It can only hold on subsets containing sufficiently few regressors. Rather than assuming such a property directly for the random sample Gram matrix, it is typically more natural to impose the corresponding RE condition on the population covariance matrix $\bSigma_x$, as a property of the DGP, and then show that it is inherited by $\hat{\bSigma}_x$ via some deviation bound such as that in Lemma \ref{lem_joint_concentration}. However, the size of the subsets is important for this transfer, since for any index set $J$, $\|\hat{\bSigma}_{x,JJ}-\bSigma_{x,JJ}\|_{\mathrm{op}} \le |J|\|\hat{\bSigma}_x-\bSigma_x\|_{\max}$. Hence, the smaller the admissible support size, the smaller the population-to-sample perturbation and the easier it is for the sample Gram matrix to inherit the population RE.\footnote{By Weyl's inequality, $\lambda_{\min}(\hat{\bSigma}_{x,JJ}) \ge \lambda_{\min}(\bSigma_{x,JJ}) - \|\hat{\bSigma}_{x,JJ}-\bSigma_{x,JJ}\|_{\mathrm{op}}$. Hence, if $\lambda_{\min}(\bSigma_{x,JJ})\ge\kappa>0$ and $\|\hat{\bSigma}_{x,JJ}-\bSigma_{x,JJ}\|_{\mathrm{op}}=o(1)$, then $\lambda_{\min}(\hat{\bSigma}_{x,JJ})$ remains bounded away from zero wpt1.} Hence, controlling the cardinality of $U$ is essential. But another BAR-specific issue here is that $U=S_\lambda\cup\hat S$ is itself random. We therefore first need to show that, with high probability, $U$ belongs to a deterministic class of subsets with controlled cardinality, and impose a preliminary RE condition uniformly over that class.\footnote{A RE condition imposed directly on $U$ would itself be data dependent, since $U$ contains the support selected by BAR from the same sample.}

The remainder of the argument proceeds in three steps. First, we derive
a preliminary deterministic upper bound on $|U|$. Second,
exploiting the BAR fixed-point equations, we further reduce this bound down to the sparsity scale $s_0\lambda^{-q_s/2}$ (cf Corollary
\ref{cor_rates}). Third, on this sufficiently smaller support scale, we transfer a population RE condition to the sample Gram matrix
using Lemma \ref{lem_EPB}. This yields the restricted curvature needed
to control $\bDelta_U$.

\begin{proposition}[Crude support bound]\label{prop_crude_bound}
Let $\hat{\bbeta}_{BAR}$ be a fixed point of Algorithm
\ref{alg:BAR_CD}. Then $|\hat S|\le \|\by\|_2^2/(4\lambda T)$. Moreover, for any fixed $C_y>0$, on the event $\mathcal E_y:=\{T^{-1}\|\by\|_2^2\le C_y\}$, \begin{align} |U| \le \bar m_T := \left\lceil 2^{-q_s}s_0\lambda^{-q_s/2} +\frac{C_y}{4\lambda} \right\rceil \asymp s_0\lambda^{-q_s/2}+\lambda^{-1}. \label{eq_m_n} \end{align} \end{proposition}
By stationarity and ergodicity under Assumption \ref{ass_phidep}, $T^{-1}\|\by\|_2^2\overset{p}{\to}\mathbb E(y_t^2)$. Hence, choosing $C_y>\mathbb E(y_t^2)$ gives $\mathbb P(\mathcal E_y)\to1$. Defining $\mathcal J_{\bar m_T}:=\{J\subseteq\{1,\ldots,Kp\}:|J|\le\bar m_T\}$, Proposition \ref{prop_crude_bound} therefore implies $U\in\mathcal J_{\bar m_T}$ with probability tending to one.\footnote{Note that Lemma \ref{lem_tailsize} is stated with the $\lesssim$ notation. The precise constant is $2^{-q_s}$, as clear from its proof given in the \autoref{app:proofs}. We use it in Proposition \ref{prop_crude_bound} just to remark that the class is deterministic.} The class $\mathcal J_{\bar m_T}$ is deterministic, so a restricted eigenvalue condition could now be imposed uniformly over it. 

The problem is that currently the obtained support bound $\bar m_T$ is too large to obtain the desired sample curvature directly from the population covariance using our Lemma \ref{lem_EPB}. In fact, we get by Lemma \ref{lem_EPB} $\norm{\hat{\bSigma}_x-\bSigma_x}_{\max}=\mathcal{O}_p(\sqrt{\lambda})$, and over a set $J$ of size $\bar m_T$, $\|\hat{\bSigma}_{x,JJ}-\bSigma_{x,JJ}\|_{\mathrm{op}}
\lesssim_p \bar m_T\sqrt{\lambda}$. So plugging in the obtained rate $\bar m_T\asymp s_0\lambda^{-q_s/2}+\lambda^{-1}$ one obtains $s_0\lambda^{(1-q_s)/2}+\lambda^{-1/2}$. Since $\lambda=C_\lambda\log(Kp)/T\to0$ under Assumption \ref{ass_phidep}, the second term diverges. Thus, at the crude support scale $\bar m_T$, the available max-norm concentration bound is not sufficient to transfer a population RE condition to the corresponding sample Gram submatrices. We therefore have to impose a preliminary sample RE condition uniformly over the deterministic class $\mathcal J_{\bar m_T}$.

\begin{assumption}[Preliminary Sample RE]
\label{ass_crudeRE}
For the deterministic sequence $\bar m_T$ in \eqref{eq_m_n} and $\mathcal J_{\bar m_T} := \left\{ J\subseteq\{1,\ldots,Kp\}: |J|\le \bar m_T \right\},$ define \begin{align} \mathcal E_{\mathrm{crudeRE}} := \left\{ \min_{J\in\mathcal J_{\bar m_T} } \lambda_{\min}(\hat{\bSigma}_{x,JJ}) \ge \kappa_{\mathrm{crude}} \right\}, \label{eq:crude_re} \end{align}
where $\kappa_{\mathrm{crude}}>0$. Assume that $\mathbb P(\mathcal E_{\mathrm{crudeRE}})\to1$.
\end{assumption}

The need for this high-level, preliminary sample RE arises because BAR determines its support through the activation threshold $\mathcal T_j(\lambda)$, which determines which coordinates remain active but does not directly constrain their number.\footnote{This differs from many iterative hard-thresholding procedures, where the sparsity level is fixed ex ante and each iteration is explicitly restricted to a prescribed number of active coordinates.} Assumption \ref{ass_crudeRE} is therefore an intermediate requirement of the present argument. It provides the preliminary curvature needed to tighten the crude bound on $|U|$ using the BAR fixed-point equations. The resulting refinement below reduces in fact the realized support union from $s_0\lambda^{-q_s/2}+\lambda^{-1}$ to only the sparsity scale $s_0\lambda^{-q_s/2}$. This is important for two reasons: first, as seen, at this scale the population-to-sample curvature transfer becomes feasible under the deviation bound of Lemma \ref{lem_EPB}.\footnote{Replacing Assumption \ref{ass_crudeRE} by primitive population conditions would require a sharper uniform concentration result for
sparse covariance submatrices directly in operator norm at the scale $\bar m_T$. Establishing such a result under the finite-moment physical dependence framework considered here is beyond the scope of the paper.} Second, as will become clear later, the support size plays a fundamental role also in order to obtain the sharp estimation and prediction error rates derived below, so regardless of the RE, the refinement from $s_0\lambda^{-q_s/2}+\lambda^{-1}$ to $s_0\lambda^{-q_s/2}$ is necessary. The starting point for this refinement is the following quadratic identity on $U$.

\begin{lemma}[Quadratic Identity on $U$]
\label{lem_shellident}
Let $\bv:=T^{-1}\bX'(\by-\bX\hat{\bbeta}_{BAR})$. On $U:=S_\lambda\cup\hat S$, the estimation error $\bDelta:=\hat{\bbeta}_{BAR}-\bbeta$ satisfies \begin{align} \bDelta_U' \hat{\bSigma}_{x,UU}\bDelta_U + \sum_{j\in U}\Delta_jv_j = \bDelta_U'T^{-1}\bX_U'\bu + \bDelta_U'\hat{\bSigma}_{x,UU^c}\bbeta_{U^c}.
\label{eq_shell_ident}
\end{align}
\end{lemma}
Equation \eqref{eq_shell_ident} provides the device for reducing the
preliminary bound on $|U|$ to the desired rate. Since $U=S_\lambda\cup\hat S$, we have $|U|\le |S_\lambda|+|F|$, where $F:=\hat S\setminus S_\lambda$ denotes the set of coordinates selected by BAR as active even though their true coefficients lie below the population threshold $\mathcal{T}(\lambda)$, i.e., the ``false positives''. The term $|S_\lambda|$ is already controlled at the desired sparsity scale by Lemma \ref{lem_tailsize}. Hence, it remains to show that $|F|$ is of the same order.  The false positives in $F$ need to be treated differently depending on the size of their true coefficients. If $|\beta_j|$ lies just below the activation threshold, selection by BAR need not imply a large estimation error, but the number of such coefficients can itself be controlled by weak sparsity. If instead $|\beta_j|$ is sufficiently below the threshold, then BAR selection, together with $|\hat\beta_j|\ge\sqrt{\lambda}$, forces a non-negligible discrepancy between $\hat\beta_j$ and $\beta_j$. The fixed-point identity \eqref{eq_shell_ident}, combined with the preliminary sample RE, can then be used to limit how many such coordinates can be selected. This motivates the following decomposition. Fix $c_0\in(0,1)$ and define \begin{align} B_\lambda &:= \{j\in S_\lambda^c: c_0\sqrt\lambda<|\beta_j|\le2\sqrt\lambda\}, & D_\lambda &:= \{j\in S_\lambda^c: |\beta_j|\le c_0\sqrt\lambda\}, \notag\\ F_B &:= F\cap B_\lambda, & F_D &:= F\cap D_\lambda.
\label{eq:buffer_deep_tail}
\end{align}
Since $F_B$ and $F_D$ form a partition of $F$, we can first control $|F_B|$ by weak sparsity.

\begin{lemma}[False positives near the threshold]
\label{lem_false_positives}
Under Assumption \ref{ass_sparsity}, \begin{align*} |F_B| \le |B_\lambda| \lesssim s_0\lambda^{-q_s/2}. \end{align*} \end{lemma} Now, since $|U|\le |S_\lambda|+|F_B|+|F_D|$ and both $|S_\lambda|$ and $|F_B|$ are already of order $s_0\lambda^{-q_s/2}$, the only component that can account for the additional $\lambda^{-1}$ term in the crude bound $\bar m_T\asymp s_0\lambda^{-q_s/2}+\lambda^{-1}$ is $F_D$. Therefore, removing the $\lambda^{-1}$ term from the bound on $|U|$ amounts to showing that $|F_D|=\mathcal O_p(s_0\lambda^{-q_s/2})$. For $j\in F_D$, we have $|\beta_j|\le c_0\sqrt{\lambda}$, while BAR selection implies $|\hat\beta_j|\ge\sqrt{\lambda}$. Hence, each such coordinate inflates the term $\sum_{j\in U}\Delta_jv_j$ on the left-hand side of \eqref{eq_shell_ident} by at least $(1-c_0)\lambda$, as formalized next.   
\begin{lemma}[Contribution of $F_D$] \label{lem_deep_tail_penalty} For every $j\in F_D$, the fixed-point identity \eqref{eq:fixed_point_identity} and the active-coordinate bound $|\hat\beta_j|\ge\sqrt\lambda$ imply \begin{align} \Delta_jv_j = \lambda\left(1-\frac{\beta_j}{\hat\beta_j}\right) \ge (1-c_0)\lambda, \qquad \sum_{j\in F_D}\Delta_jv_j \ge (1-c_0)\lambda|F_D|.
\label{eq:deep_tail_coercivity}
\end{align}
\end{lemma}
If $d=|F_D|$ such coordinates are selected, their contribution to $\sum_{j\in U}\Delta_jv_j$ is at least $(1-c_0)\lambda d$. Equation \eqref{eq_shell_ident} then limits how large $d$ can be. In particular, on $U\in\mathcal J_{\bar m_T}$, Assumption \ref{ass_crudeRE} gives the lower bound $\bDelta_U'\hat{\bSigma}_{x,UU}\bDelta_U \ge\kappa_{\mathrm{crude}}\|\bDelta_U\|_2^2$. The empirical-noise term $\bDelta_U'T^{-1}\bX_U'\bu$ is controlled on $\mathcal E_\chi$ by Lemma \ref{lem_EPB}, while the interaction $\bDelta_U'\hat{\bSigma}_{x,UU^c}\bbeta_{U^c}$ is controlled using the Gram-matrix deviation in the same lemma, combined with an upper eigenvalue bound on $\bSigma_x$, and the bounds on $\|\bbeta_{S_\lambda^c}\|_r$, $r\in\{1,2\}$. Finally, the remaining terms that can contribute inflating $|F_D|$, namely $S_\lambda$ and $F_B$, are already controlled by their cardinalities. Balancing these terms against $(1-c_0)\lambda d$ yields the following refinement. 

\begin{proposition}[Support-union dimension refinement]
\label{prop_shell_dimension}
Assume $\lambda_{\max}(\bSigma_x)\le C_{\max}$ and let $\lambda=C_\lambda\log(Kp)/T$, $c_u:=C_\chi/\sqrt{C_\lambda}$, $a_u:=c_u^2/\kappa_{\mathrm{crude}}$, and $\mathcal G:=\mathcal E_y\cap\mathcal E_{\mathrm{crudeRE}} \cap\mathcal E_\chi$. If $a_u<1-c_0$ and $s_0\lambda^{(1-q_s)/2}=o(1)$, then on $\mathcal G$, \begin{align} |F_D| \le C_1\bigl(|S_\lambda|+|F_B|\bigr) + C_2\lambda^{-1} \left\{ \|\bbeta_{S_\lambda^c}\|_2^2 + \sqrt{\lambda}\, \|\bbeta_{S_\lambda^c}\|_1^2 \right\}. \label{eq:shell_dimension_refinement}
\end{align}
 
Consequently, there exists a fixed constant $C_U>0$ such that, for
all sufficiently large $T$,
\begin{align}
|U|
\le
C_U s_0\lambda^{-q_s/2}
\qquad\text{on }\mathcal G.
\label{eq:refined_support_bound}
\end{align}
Hence, if $\mathbb P(\mathcal G)\to1$,
$|U|=\mathcal O_p(s_0\lambda^{-q_s/2})$.\end{proposition}

The bound in \eqref{eq:shell_dimension_refinement} is of the desired order. Indeed, $|S_\lambda|+|F_B| \lesssim s_0\lambda^{-q_s/2}$, while $\lambda^{-1}\|\bbeta_{S_\lambda^c}\|_2^2 \lesssim s_0\lambda^{-q_s/2}$. The remaining term is of smaller order under $s_0\lambda^{(1-q_s)/2}=o(1)$, since $s_0\lambda^{(1-q_s)/2} =(s_0\lambda^{-q_s/2})\sqrt{\lambda}=o(1)$. Hence $|F_D|=\mathcal O_p(s_0\lambda^{-q_s/2})$ and, together with the bounds for $S_\lambda$ and $F_B$, $|U|=\mathcal O_p(s_0\lambda^{-q_s/2})$.

Having shrunk down the cardinality of $U$ from $\bar m_T \asymp s_0\lambda^{-q_s/2}+\lambda^{-1}$ to $m_T^\sharp
:=
\left\lceil C_U s_0\lambda^{-q_s/2}\right\rceil$, we can use Lemma \ref{lem_EPB} to transfer a population RE condition to the sample Gram matrix on this refined support scale. Thus, although the high-level
sample condition in Assumption \ref{ass_crudeRE} is needed for the
preliminary support-refinement step, the curvature used in the final
error analysis can instead be derived from the more natural population
RE condition in \eqref{eq:population_shell_re}.
\begin{lemma}[RE on the BAR support union]
\label{lem_refined_RE}
Assume
that
\begin{align}
\min_{\substack{J\subseteq\{1,\ldots,Kp\}\\ |J|\le m_T^\sharp}}
\lambda_{\min}(\bSigma_{x,JJ})
\ge \kappa>0.
\label{eq:population_shell_re}
\end{align}
Under $s_0\lambda^{(1-q_s)/2}=o(1)$, for all sufficiently large
$T$, on $\{|U|\le m_T^\sharp\}\cap\mathcal E_\chi$,
$\lambda_{\min}(\hat{\bSigma}_{x,UU})\ge\kappa/2$.
\end{lemma}
Indeed, on $\{|U|\le m_T^\sharp\}\cap\mathcal E_\chi$, Lemma
\ref{lem_joint_concentration} gives
\begin{align*}
\|\hat{\bSigma}_{x,UU}-\bSigma_{x,UU}\|_{\mathrm{op}}
&\le
C_\chi m_T^\sharp\sqrt{\frac{\log(Kp)}{T}}
=
\mathcal O\!\left(s_0\lambda^{(1-q_s)/2}\right)
=o(1),
\end{align*}
where the last step follows from
$s_0\lambda^{(1-q_s)/2}=o(1)$. Weyl's inequality then yields the
claimed sample RE. By Proposition \ref{prop_shell_dimension} and Lemma
\ref{lem_joint_concentration}, the event
$\{|U|\le m_T^\sharp\}\cap\mathcal E_\chi$ has probability tending to
one.

We can now return to the cross-design term that motivated this subsection. Its difficulty came from the fact that it depends on the unknown estimation error $\bDelta$. The support-union representation separates this error into the unknown component $\bDelta_U$ and the controlled component $\bDelta_{U^c}=-\bbeta_{U^c}$. In the quadratic identity \eqref{eq_shell_ident}, the former is handled through the restricted curvature of $\hat{\bSigma}_{x,UU}$, while the latter is controlled by the weak-sparsity bounds together with the Gram-matrix deviation. This resolves the cross-design difficulty without requiring a direct coordinatewise bound on the original term.
%%%%%%%%%%%%%%%%%%%

With the RE condition verified on the refined support scale $m_T^\sharp$, we can now derive the high-probability error bounds for the BAR estimator. We return to the quadratic inequality used in Proposition \ref{prop_shell_dimension}, substitute the sample curvature bound $\lambda_{\min}(\hat{\bSigma}_{x,UU})\ge\kappa/2$, and solve for $\|\bDelta_U\|_2$. The full $\ell_2$ error then follows from $\bDelta=\bDelta_U-\bbeta_{U^c}$ together with the bounds on $\|\bbeta_{U^c}\|_2$. The $\ell_1$ rate is obtained similarly using $\|\bDelta_U\|_1\le\sqrt{|U|}\|\bDelta_U\|_2$, while the prediction error follows from the same quadratic inequality.

\begin{theorem}[Estimation and Prediction Error Rates]\label{thm_estimation_rates}
Suppose the conditions of Proposition \ref{prop_shell_dimension} and Lemma \ref{lem_refined_RE} hold and $\lambda\asymp C_\lambda\log(Kp)/T$. On the event $\mathcal E_T:=\mathcal G\cap\{|U|\le m_T^\sharp\}$, whose probability tends to one, \begin{align}\label{eq:bar_error_rates} \|\hat{\bbeta}_{BAR}-\bbeta\|_2 &=\mathcal O_p\left(\sqrt{s_0\lambda^{1-q_s/2}}\right),& \|\hat{\bbeta}_{BAR}-\bbeta\|_1 &=\mathcal O_p\left(s_0\lambda^{\frac{1-q_s}{2}}\right),\\ T^{-1}\|\bX(\hat{\bbeta}_{BAR}-\bbeta)\|_2^2 &=\mathcal O_p\left(s_0\lambda^{1-q_s/2}\right).&& \end{align}
\end{theorem}

\begin{corollary}\label{cor_asymp}
Under the conditions of Theorem \ref{thm_estimation_rates}, the prediction, $\ell_2$, and $\ell_1$ errors all vanish if \begin{align} s_0 \left\{ \frac{\log(Kp)}{T} \right\}^{(1-q_s)/2} \to0. \label{eq_sparsityrate_common} \end{align} In particular, bounded $s_0$ suffices.
\end{corollary}

Condition \eqref{eq_sparsityrate_common} coincides with
$s_0\lambda^{(1-q_s)/2}=o(1)$ from Proposition \ref{prop_shell_dimension}, which was already used for the
support refinement and the population-to-sample RE transfer. It is also
the condition required for the $\ell_1$ error in Theorem
\ref{thm_estimation_rates} to vanish, while the prediction and $\ell_2$
rates alone allow weaker sparsity growth. Under Assumption
\ref{ass_phidep}, $Kp=O(T^\nu)$ for some fixed $\nu>0$, so a sufficient
growth condition is
$s_0=o\{(T/\log T)^{(1-q_s)/2}\}$.

\subsection{Information-criterion tuning of the BAR penalty}
\label{sec_lambda_ic}

The theory considers the deterministic penalty sequence $\lambda=C_{\lambda}\log(Kp)/T$. The multiplier $C_{\lambda}$ must be sufficiently large relative to the constants entering the concentration and preliminary curvature bounds; in particular, Proposition \ref{prop_shell_dimension} requires $C_{\lambda} > \frac{C_\chi^2} {\kappa_{\mathrm{crude}}(1-c_0)}$.  This condition is not directly feasible, however, since $C_\chi$ and $\kappa_{\mathrm{crude}}$ are unknown without strong parametric assumptions on the innovation distribution, the dependence and design structure. Algorithm \ref{alg:BAR_path} therefore selects the penalty from a data-driven grid using an information criterion. The Ridge initializer parameter $\xi$ is selected by BIC, whereas the sparsity-inducing penalty
$\lambda$ is selected along the BAR path by an information criterion. The two tuning parameters play distinct roles but are not independent:
$\xi$ determines the initial Ridge estimate $\hat{\bbeta}^{(0)}$ and
thereby affects the scale of the data-driven BAR penalty path, whereas
$\lambda$ governs thresholding and sparsification.

For each candidate $\lambda$, let
$RSS(\lambda):=\|\by-\bX\hat{\bbeta}_{BAR}(\lambda)\|_2^2$ and
$\hat S(\lambda):=\supp\{\hat{\bbeta}_{BAR}(\lambda)\}$. For the
sparsity-inducing estimators considered below, we use the generic
criterion
\begin{align}
IC_\tau(\lambda)
=
\log\{RSS(\lambda)/T\}
+
\frac{\tau_T}{T}|\hat S(\lambda)|.
\label{eq_generic_ic}
\end{align}
The usual choices are $\tau_T=2$ for AIC,
$2\log\log T$ for HQ, $\log T$ for BIC, $\log T+1$ for CAIC, and
$\log T+2\gamma\log(Kp)$ for EBIC, with $\gamma\in[0,1]$.
The BIC used to select the Ridge initializer $\xi$ is computed separately
using the corresponding Ridge effective degrees of freedom, rather
than the support cardinality in \eqref{eq_generic_ic}.

Under the polynomial dimensional regime in Assumption
\ref{ass_phidep}(iv), $\log(Kp)=O(\log T)$, so BIC and EBIC have
complexity weights of the same asymptotic order. EBIC nevertheless
adds the dimension-dependent term $2\gamma\log(Kp)$ and can therefore
impose substantially stronger penalization when the number of
candidate regressors is large. We use BIC for the Ridge initializer and EBIC
for the sparsity-inducing BAR penalty as a practical calibration
choice. Accordingly, the theoretical results above apply to the
deterministic penalty sequence
$\lambda=C_{\lambda}\log(Kp)/T$ and are not intended as a formal
justification of the EBIC-selected penalty. In particular, we do not
claim that the data-driven $\hat\lambda$ automatically satisfies the
sufficiently-large-multiplier condition required by the theoretical
results.

Nonetheless, the tuning behavior in the simulations provides further motivation for
this choice. Table \ref{tab:lambda_pos_median} reports the median
position of the selected penalty in the 100-point grid, ordered from
the largest $\lambda$ at position 1 to the smallest $\lambda$ at
position 100. Hence, larger positions correspond to weaker
regularization.

\begin{table}[H]
\centering
\small
\caption{Median position of the selected $\lambda$ in the 100-point
grid}
\label{tab:lambda_pos_median}
\begin{tabular}{llccccc}
\toprule
& Criterion
& BAR
& \makecell{aLasso\\(Ridge)}
& \makecell{aLasso\\(Lasso)}
& Lasso
& SCAD \\
\midrule
\multirow{2}{*}{$K\le T$}
& BIC  & 75 & 22 & 56  & 13  & 20 \\
& EBIC & 62 & 20 & 55  & 13  & 17 \\
\cmidrule{1-7}
\multirow{2}{*}{$K>T$}
& BIC  & 96 & 37 & 100 & 100 & 99 \\
& EBIC & 49 & 6  & 1   & 1   & 16  \\
\bottomrule
\end{tabular}
\end{table}

When $K\le T$, BIC and EBIC generally select similar regions of each
estimator's tuning path. When $K>T$, however, BIC frequently selects
penalties at or near the weakest-regularization boundary. For BAR, for
example, the median selected position moves from 96 under BIC to 49
under EBIC, while analogous shifts toward stronger regularization are
visible for the competing sparse estimators. Thus, EBIC avoids the
systematic tendency of BIC to select the weakest penalties in the
high-dimensional designs considered here. This evidence does not
establish selection consistency or verify the sufficiently-large-
multiplier condition used in the theory, but it provides a 
finite-sample motivation for using EBIC to tune the sparsity-inducing
penalty. 
%%%%%%%%%%%%%%%%%%%%% SIMULATIONS %%%%%%%%%%%%%%%%%%%%%%%
\section{Monte Carlo simulations}\label{sec:simulations}

% We assess the finite-sample performance of BAR through a simulation study based on the $K$-dimensional VAR(1) \begin{align} \by_t^{(m)}=\bA^{(m)}\by_{t-1}^{(m)}+\bu_t^{(m)},  \qquad \bA^{(m)} =\bV^{-1}\bD^{(m)}\bV, \label{eq:sim_dgp} \end{align} where $m=1,\ldots,N_{\mathrm{MC}}$ indexes the Monte Carlo replication.\footnote{\add{A replication package is available at XXXXXX}} Specifically, $\bV=\bI_{K/r}\otimes\bnu$, with $[\bnu]_{ij}=\cos(2\pi ij/r)+\sin(2\pi ij/r)$, while $\bD^{(m)} = \bI_{K/r}\otimes \operatorname{diag}(\bs^{(m)}\odot\bdelta)$, where $\bdelta=(\delta_1,\ldots,\delta_r)'$ contains equally spaced values in $[0.25,0.75]$ and the entries of $\bs^{(m)}$ are independent Rademacher variables redrawn in each replication.
We assess the finite-sample performance of BAR through a simulation study based on the $K$-dimensional VAR(1) \begin{align} \by_t=\bA\by_{t-1}+\bu_t,  \qquad \bA =\bV^{-1}\bD\bV, \label{eq:sim_dgp} \end{align} where, for a block of size $r$, $\bV=\bI_{K/r}\otimes\bnu$, with $[\bnu]_{ij}=\cos(2\pi ij/r)+\sin(2\pi ij/r)$, while $\bD = \bI_{K/r}\otimes \operatorname{diag}(\bs\odot\bdelta)$, where $\bdelta=(\delta_1,\ldots,\delta_r)'$ contains equally spaced values in $[0.25,0.75]$ and $\bs=(s_1,\ldots,s_r)'$ is a deterministic alternating-sign vector ($s_j=(-1)^{j-1}$).\footnote{A replication package is available at \url{https://github.com/orestprifti/TS-BAR}.} When $r=1$, we set $\delta_1=0.5$. The resulting $\bA$ is block diagonal, and each equation contains exactly $r$ nonzero autoregressive coefficients. Thus $r=1$ yields a diagonal system, whereas larger values of $r$ generate progressively denser within-block dynamics. Each replication is generated with a burn-in of 50 observations, which is discarded. The following $T$ observations are used for estimation, while the subsequent observation $T+1$ is reserved for one-step-ahead forecast evaluation. The innovations are Gaussian, $\bu_t\overset{\mathrm{iid}}{\sim}\mathcal{N}_K(\bzero,\bSigma_u)$, with Toeplitz covariance matrix $[\bSigma_u]_{ij}=\rho^{|i-j|}$. We consider $\rho\in\{0,0.5\}$, where $\rho=0$ corresponds to cross-sectionally independent innovations and $\rho=0.5$ introduces contemporaneous dependence. The simulation design covers $K\in\{20,40,80,160,320\}$, $T\in\{50,100,500\}$, $r\in\{1,2,5,10\}$, and $\rho\in\{0,0.5\}$. We compare BAR with adaptive Lasso initialized by Ridge (aLASSO(ridge)), adaptive Lasso initialized by Lasso (aLASSO(lasso)), Lasso, and SCAD. The sparsity-inducing penalties are selected by EBIC with $\gamma=0.5$. The adaptive-Lasso weights are constructed as the inverse absolute value of the corresponding initial estimates, while the SCAD shape parameter is set to $3.7$. All estimators use the same centering and standardization convention. 

Performance is evaluated through the five criteria summarized in
Table \ref{tab:MC_measures}. Let $\mathcal S
=\{(i,j):|A_{ij}|>0\}$
denote the true support and
$\widehat{\mathcal S}$ its estimate.

\begin{table}[H]
\centering
\caption{Monte Carlo performance measures.}
\label{tab:MC_measures}
\small
\renewcommand{\arraystretch}{1.15}
\begin{tabular}{@{}l@{\hspace{2em}}l@{}}
\toprule
Measure & Definition \\
\midrule

RMSFE &
$\left[
N_{\mathrm{MC}}^{-1}\sum_{m=1}^{N_{\mathrm{MC}}}
K^{-1}\|\widehat{\by}_{T+1,m}-\by_{T+1,m}\|_2^2
\right]^{1/2}$ \\

RMSE &
$\left[
N_{\mathrm{MC}}^{-1}\sum_{m=1}^{N_{\mathrm{MC}}}
\|\widehat{\bA}_{m}-\bA\|_F^2
\right]^{1/2}$ \\

Correct sparsity &
$N_{\mathrm{MC}}^{-1}\sum_{m=1}^{N_{\mathrm{MC}}}
\mathds{1}\{\widehat{\mathcal S}_{m}=\mathcal S\}$ \\

True model included &
$N_{\mathrm{MC}}^{-1}\sum_{m=1}^{N_{\mathrm{MC}}}
\mathds{1}\{\mathcal S\subseteq\widehat{\mathcal S}_{m}\}$ \\

Relevant coefficients selected &
$N_{\mathrm{MC}}^{-1}\sum_{m=1}^{N_{\mathrm{MC}}}
\frac{
|\mathcal S\cap\widehat{\mathcal S}_{m}|
}{
|\mathcal S|
}$ \\

\bottomrule
\end{tabular}
\end{table}

\begin{figure}[!t]
\centering

\newcommand{\mcbar}[4]{%
    \filldraw[
        fill=#1,
        draw=#1!65!black,
        line width=0.2pt
    ]
    (0,#2) rectangle (#3,#2+0.48);
    \node[
        anchor=west,
        font=\tiny
    ] at (#3+1,#2+0.24) {#4};
}

\begin{tikzpicture}[x=0.060cm,y=0.34cm,font=\scriptsize]

% Vertical grid and horizontal-axis labels
\foreach \x in {0,20,40,60,80,100}{
    \draw[gray!30,dotted] (\x,-3.25) -- (\x,18.75);
    \node[anchor=north] at (\x,-3.35) {\x};
}

\draw[black!65] (0,-3.25) -- (100,-3.25);
\node[font=\small] at (50,-5.2) {Average win rate (\%)};

% Criterion labels
\node[anchor=east,align=right] at (-3,16.95) {RMSFE};
\node[anchor=east,align=right] at (-3,12.45) {RMSE};
\node[anchor=east,align=right] at (-3,7.95) {Correct sparsity};
\node[anchor=east,align=right] at (-3,3.45) {True model included};
\node[anchor=east,align=right] at (-3,-1.05)
    {Relevant coefficients\\selected};

% RMSFE
\mcbar{RoyalBlue}{18.00}{77.5}{77.5}
\mcbar{Orange}{17.35}{47.5}{47.5}
\mcbar{ForestGreen}{16.70}{46.7}{46.7}
\mcbar{BrickRed}{16.05}{7.5}{7.5}
\mcbar{Purple}{15.40}{22.5}{22.5}

% RMSE
\mcbar{RoyalBlue}{13.50}{42.5}{42.5}
\mcbar{Orange}{12.85}{28.3}{28.3}
\mcbar{ForestGreen}{12.20}{17.5}{17.5}
\mcbar{BrickRed}{11.55}{4.2}{4.2}
\mcbar{Purple}{10.90}{8.3}{8.3}

% Correct sparsity
\mcbar{RoyalBlue}{9.00}{5.8}{5.8}
\mcbar{Orange}{8.35}{0.8}{0.8}
% aLasso(lasso) is zero and is therefore not drawn
% Lasso is zero and is therefore not drawn
\mcbar{Purple}{6.40}{8.3}{8.3}

% True model included
\mcbar{RoyalBlue}{4.50}{5.8}{5.8}
\mcbar{Orange}{3.85}{6.7}{6.7}
\mcbar{ForestGreen}{3.20}{8.3}{8.3}
\mcbar{BrickRed}{2.55}{12.5}{12.5}
\mcbar{Purple}{1.90}{9.2}{9.2}

% Relevant coefficients selected
\mcbar{RoyalBlue}{0.00}{72.5}{72.5}
\mcbar{Orange}{-0.65}{10.0}{10.0}
\mcbar{ForestGreen}{-1.30}{11.7}{11.7}
\mcbar{BrickRed}{-1.95}{30.0}{30.0}
\mcbar{Purple}{-2.60}{15.8}{15.8}

\end{tikzpicture}

\vspace{0.4em}

\begin{tabular}{ccccc}
\textcolor{RoyalBlue}{\rule{1.3ex}{1.3ex}} BAR
&
\textcolor{Orange}{\rule{1.3ex}{1.3ex}} aLASSO(ridge)
&
\textcolor{ForestGreen}{\rule{1.3ex}{1.3ex}} aLASSO(lasso)
&
\textcolor{BrickRed}{\rule{1.3ex}{1.3ex}} Lasso
&
\textcolor{Purple}{\rule{1.3ex}{1.3ex}} SCAD
\end{tabular}

\caption{\footnotesize Average Monte Carlo win rates across the six combinations of
$T\in\{50,100,500\}$ and $\rho\in\{0,0.5\}$. Each underlying win rate
is aggregated over $(K,r)$. For RMSFE and RMSE, a win corresponds to
the smallest value; for the remaining criteria, it corresponds to the
largest value. Since ties are counted as wins for all tied methods,
the win rates need not sum to $100\%$.}
\label{fig:MC_average_wins}
\end{figure}
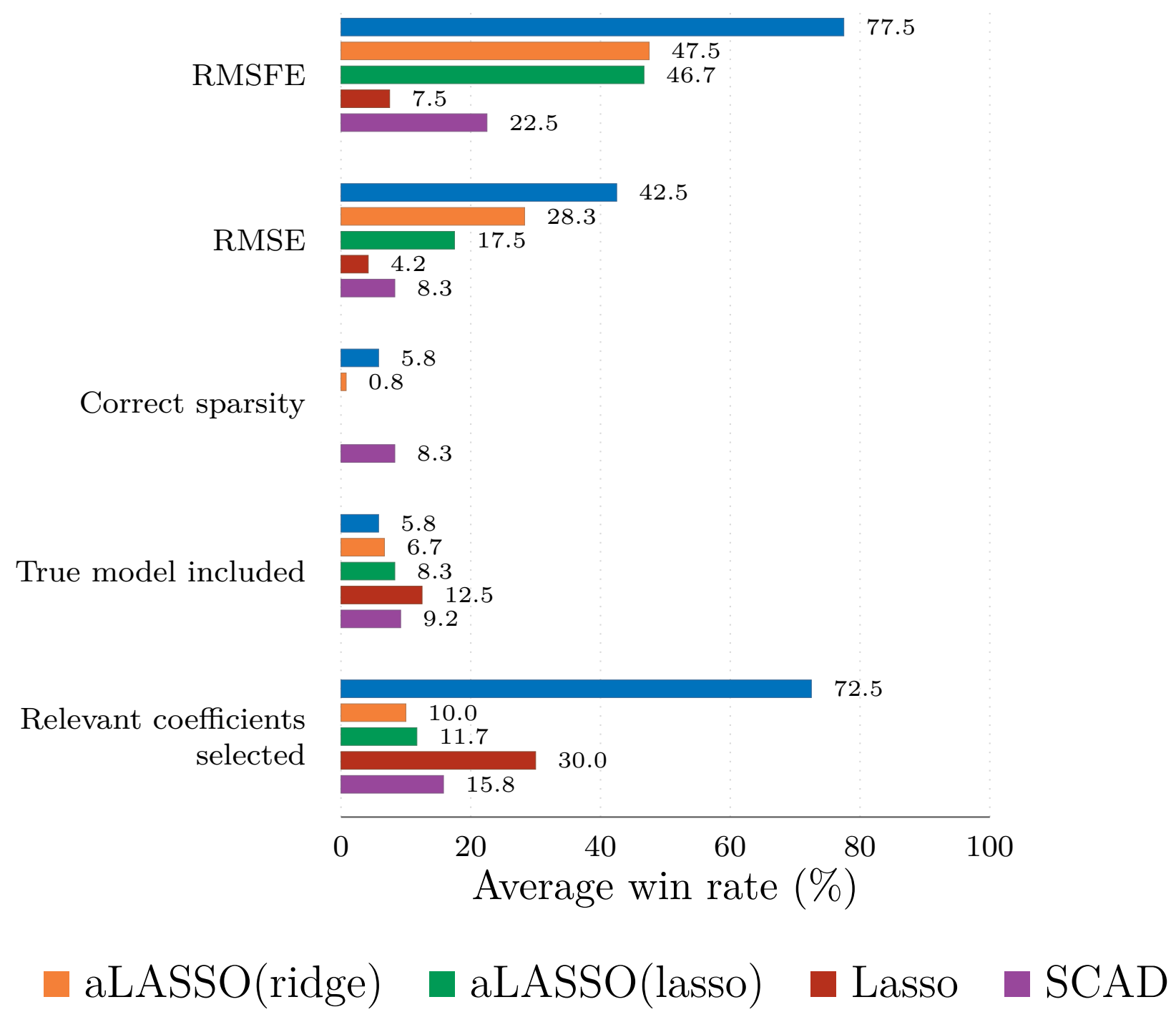

For each $(T,\rho)$, the performance measures are first computed over the $N_{\mathrm{MC}}=1000$ replications for each of the 20 combinations of $(K,r)$. A win is assigned to the estimator attaining the best value after rounding the corresponding measure to three decimal places. Ties are counted as wins for all tied estimators, except when all methods attain zero for Correct sparsity or True model included, in which case no win is assigned. The win rates are then computed across the 20 $(K,r)$ designs. Figure \ref{fig:MC_average_wins} reports their simple averages over the six combinations of $T\in\{50,100,500\}$ and $\rho\in\{0,0.5\}$.

As clear, BAR achieves the highest
average win rate by a wide margin for forecasting accuracy, estimation
accuracy, and the share of relevant variables selected. As expected, correct sparsity remains difficult for all methods, with SCAD attaining the
highest average win rate for this criterion. True-model inclusion is
more mixed, with Lasso performing best on average. The disaggregated
results in the \autoref{app:sim} show how these patterns vary with the sample size and contemporaneous innovation dependence.
%%%%%%%%%%%%%%%%%%%%% EMPIRICAL APPLICATION %%%%%%%%%%%%%%%%%%%%%%% 
\section{Empirical application}
We examine whether the BAR improves forecasts of U.S. inflation using a large set of economic indicators.
Accurate inflation forecasts are crucial for policymakers, financial institutions, firms, and households. Central banks rely on them to conduct monetary policy and anchor inflation expectations, while financial markets incorporate them into interest rates, asset prices, and investment decisions. Inflation expectations also shape firms’ and households’ employment, investment, and consumption choices.
We use the FRED-MD database developed by \cite{McCrackenNg2016}, which contains 126 monthly U.S. macroeconomic series spanning 1959--2025.\footnote{The dataset is publicly available and regularly updated on Michael McCracken’s webpage at the Federal Reserve Bank of St. Louis website: \url{https://www.stlouisfed.org/research/economists/mccracken/fred-databases}. We use the June 2026 vintage and apply the variable-specific transformations reported on the same webpage.} 
To obtain the longest balanced sample, we exclude five series with substantial missing values and retain data from January 1960 to December 2025, yielding 792 monthly observations.\footnote{Including the series ACOGNO, ANDENOx, TWEXAFEGSMTHx, UMCSENTx, and VIXCLSx would restrict the sample to 1992--2025.}
The target variable is the all-items Consumer Price Index (CPIAUCSL).
We evaluate forecast accuracy over three out-of-sample periods, each corresponding to a distinct economic environment:

\begin{itemize}
    \item 1990--2019: the full pre-COVID out-of-sample period;
    \item 2020--2021: the COVID-19 pandemic, a major exogenous shock;
    \item 2022--2025: the post-pandemic period, encompassing the inflation surge and subsequent disinflation during the monetary tightening cycle.
\end{itemize}

For forecasting, we apply BAR to the single-equation regression representation underlying the VAR analysis, allowing a forecasting-specific lag structure with four lags of inflation and the current value of each additional predictor:
\begin{equation}\label{eq:forecast_macro}
    \hat{y}_{t+h}=\sum_{i=1}^{4}\hat{\phi}_{i} y_{t-i+1} + \sum_{j=1}^{K-1} \hat{\theta}_{j} x_{j,t},
\end{equation}
where \(h\) denotes forecast horizon and \(K-1\) is the number of additional predictors. We consider four forecast horizons, \(h\)\(\in\)\{1,3,6,12\}, covering a full year.
Following \citet{MedeirosRF2021}, the models are estimated over rolling windows. We use a window of $W=360$ observations for forecasts produced over 1990--2000 and $W=492$ observations from 2001 onward.
Accordingly, the effective estimation sample is $R_{h,p}=W-h-p+1$, where $p=4$ denotes the maximum lag order.
For each horizon, all models are estimated using the direct forecasting approach, with the regularization parameter \(\lambda\) selected by minimizing the EBIC.
All predictors are standardized to have zero mean and unit variance within each rolling estimation window. 
Forecast accuracy is evaluated using two measures:\footnote{In Table \ref{tab:MD_inflation_lags1_roll360}, the terms \lq aLASSO(ridge)\rq\ and \lq aLASSO(lasso)\rq\ refer to the Adaptive LASSO models in which the initial weights are obtained from Ridge and LASSO regressions, respectively.}

\begin{itemize}
    \item Root Mean Squared Forecast Error (RMSFE): 
    \(\sqrt{\frac{1}{T} \sum_{t=1}^{T} (\hat{e}_{t+h\mid t})^2}\),
    \item Mean Absolute Forecast Error (MAFE): 
    \(\frac{1}{T} \sum_{t=1}^{T} \left| \hat{e}_{t+h\mid t} \right|\),
\end{itemize}
where \(\hat{e}_{t+h\mid t}\)=\(\hat{y}_{t+h\mid t}-y_{t+h}\) denotes forecast error, and \(T\) is the number of out-of-sample forecasts.
Table \ref{tab:MD_inflation_lags1_roll360} reports RMSFE and MAFE ratios relative to the AR(1) benchmark, so that values below one indicate superior forecast accuracy.
The colored cells identify the Set of Superior Models (SSM) obtained from the 50\% Model Confidence Set (MCS) procedure of \cite{Hansen2011}. Gray cells identify models included in the SSM under squared-error loss, whereas blue cells indicate inclusion under absolute-error loss.
During the pre-COVID period (1990--2019), BAR dominates at the one-month horizon: it achieves the lowest RMSFE and MAFE ratios and is the only model included in the SSM. At longer horizons, differences across models are relatively small. During the COVID period (2020--2021), BAR generally underperforms, except at \(h\)=12, where it ranks first. In the post-COVID period (2022--2025), its performance is broadly comparable to that of the competing models.

% ---- Table ----
\begin{table}[htbp]
\small
\centering
\caption{RMSFE and MAFE ratios relative to the AR(1).}
\label{tab:MD_inflation_lags1_roll360}
\resizebox{\textwidth}{!}{%
\begin{tabular}{cccccccccc}
   \toprule
\multirow{2}{*}{period} & \multirow{2}{*}{model} & \multicolumn{2}{c}{h=1} & \multicolumn{2}{c}{h=3} & \multicolumn{2}{c}{h=6} & \multicolumn{2}{c}{h=12} \\
\cmidrule(lr){3-4} \cmidrule(lr){5-6} \cmidrule(lr){7-8} \cmidrule(lr){9-10}
{} & {} & RMSFE & MAFE & RMSFE & MAFE & RMSFE & MAFE & RMSFE & MAFE \\ 
 \midrule
\multirow{6}{*}{1990-2019} & BAR & \cellcolor{gray!30}{\underline{\textbf{0.86}}} & \cellcolor{cyan!10}{\underline{\textbf{0.90}}} & \cellcolor{gray!30}{\textbf{0.97}} & \cellcolor{cyan!10}{\textbf{0.95}} & \textbf{0.95} & 0.94 & 1.02 & 1.02 \\ 
  {} & aLASSO(ridge) & 0.93 & 0.94 & \textbf{0.97} & \textbf{0.95} & \textbf{0.95} & \textbf{0.93} & \textbf{1.01} & \underline{\textbf{1.01}} \\ 
  {} & aLASSO(lasso) & 0.91 & 0.93 & \textbf{0.97} & \textbf{0.95} & \textbf{0.95} & \textbf{0.93} & \textbf{1.01} & 1.02 \\ 
  {} & LASSO & 0.94 & 0.95 & \cellcolor{gray!30}{\textbf{0.97}} & \cellcolor{cyan!10}{\textbf{0.95}} & \cellcolor{gray!30}{\textbf{0.95}} & \cellcolor{cyan!10}{\textbf{0.93}} & \cellcolor{gray!30}{\textbf{1.01}} & \cellcolor{cyan!10}{1.02} \\ 
  {} & SCAD & 0.92 & 0.93 & \cellcolor{gray!30}{\textbf{0.97}} & \cellcolor{cyan!10}{\textbf{0.95}} & \cellcolor{gray!30}{\textbf{0.95}} & \cellcolor{cyan!10}{\textbf{0.93}} & \cellcolor{gray!30}{\textbf{1.01}} & \cellcolor{cyan!10}{1.02} \\ 
   \midrule
\multirow{6}{*}{2020-2021} & BAR & 1.10 & 0.97 & \textbf{1.05} & \textbf{1.07} & 1.21 & 1.17 & \cellcolor{gray!30}{\underline{\textbf{0.98}}} & \cellcolor{cyan!10}{\underline{\textbf{0.98}}} \\ 
  {} & aLASSO(ridge) & 1.07 & 0.97 & \textbf{1.05} & \textbf{1.07} & \textbf{0.97} & \textbf{0.95} & 1.03 & 1.02 \\ 
  {} & aLASSO(lasso) & 1.29 & 0.93 & \textbf{1.05} & \textbf{1.07} & \textbf{0.97} & \textbf{0.95} & 1.03 & 1.02 \\ 
  {} & LASSO & \cellcolor{gray!30}{\underline{\textbf{0.84}}} & \cellcolor{cyan!10}{\underline{\textbf{0.77}}} & \textbf{1.05} & \textbf{1.07} & \cellcolor{gray!30}{\textbf{0.97}} & \cellcolor{cyan!10}{\textbf{0.95}} & 1.03 & 1.02 \\ 
  {} & SCAD & 1.22 & 0.96 & \textbf{1.05} & \textbf{1.07} & \cellcolor{gray!30}{\textbf{0.97}} & \cellcolor{cyan!10}{\textbf{0.95}} & 1.03 & 1.02 \\ 
   \midrule
\multirow{6}{*}{2022-2025} & BAR & 0.91 & 0.93 & \cellcolor{gray!30}{0.95} & \cellcolor{cyan!10}{\textbf{0.98}} & \textbf{1.04} & \textbf{1.04} & \textbf{0.98} & \cellcolor{cyan!10}{\underline{\textbf{0.99}}} \\ 
  {} & aLASSO(ridge) & 0.93 & 0.95 & 0.95 & \textbf{0.98} & \textbf{1.04} & \textbf{1.04} & \textbf{0.98} & 1.00 \\ 
  {} & aLASSO(lasso) & \underline{\textbf{0.87}} & \underline{\textbf{0.92}} & 0.95 & \textbf{0.98} & \textbf{1.04} & \textbf{1.04} & \textbf{0.98} & 1.00 \\ 
  {} & LASSO & 0.92 & 0.94 & 0.95 & \textbf{0.98} & \textbf{1.04} & \textbf{1.04} & \cellcolor{gray!30}{\textbf{0.98}} & \cellcolor{cyan!10}{1.00} \\ 
  {} & SCAD & 0.91 & 0.93 & 0.95 & \textbf{0.98} & \textbf{1.04} & \textbf{1.04} & \cellcolor{gray!30}{\textbf{0.98}} & \cellcolor{cyan!10}{1.00} \\ 
   \bottomrule
\multicolumn{10}{l}{\begin{minipage}{\dimexpr\columnwidth+13\tabcolsep}
\footnotesize\textit{Notes}: the RMSFE and the MAFE are relative to the AR(1) benchmark. Bold and underlined entries denote the best-performing model(s) for each forecast horizon and period. Grey cells indicate models belonging to the Set of Superior Models (SSM) under the 50\% MCS procedure using the $T_{\text{max}}$ statistic with squared errors, whereas blue cells indicate models included in the SSM when absolute errors are used.
\end{minipage}
}\end{tabular}
} % end resizebox
\end{table}

% ---- Table ----
  
\begin{figure}[htbp] 
    \centering
    \caption{CSSFED relative to the AR(1).}
    \includegraphics[width=0.95\textwidth]{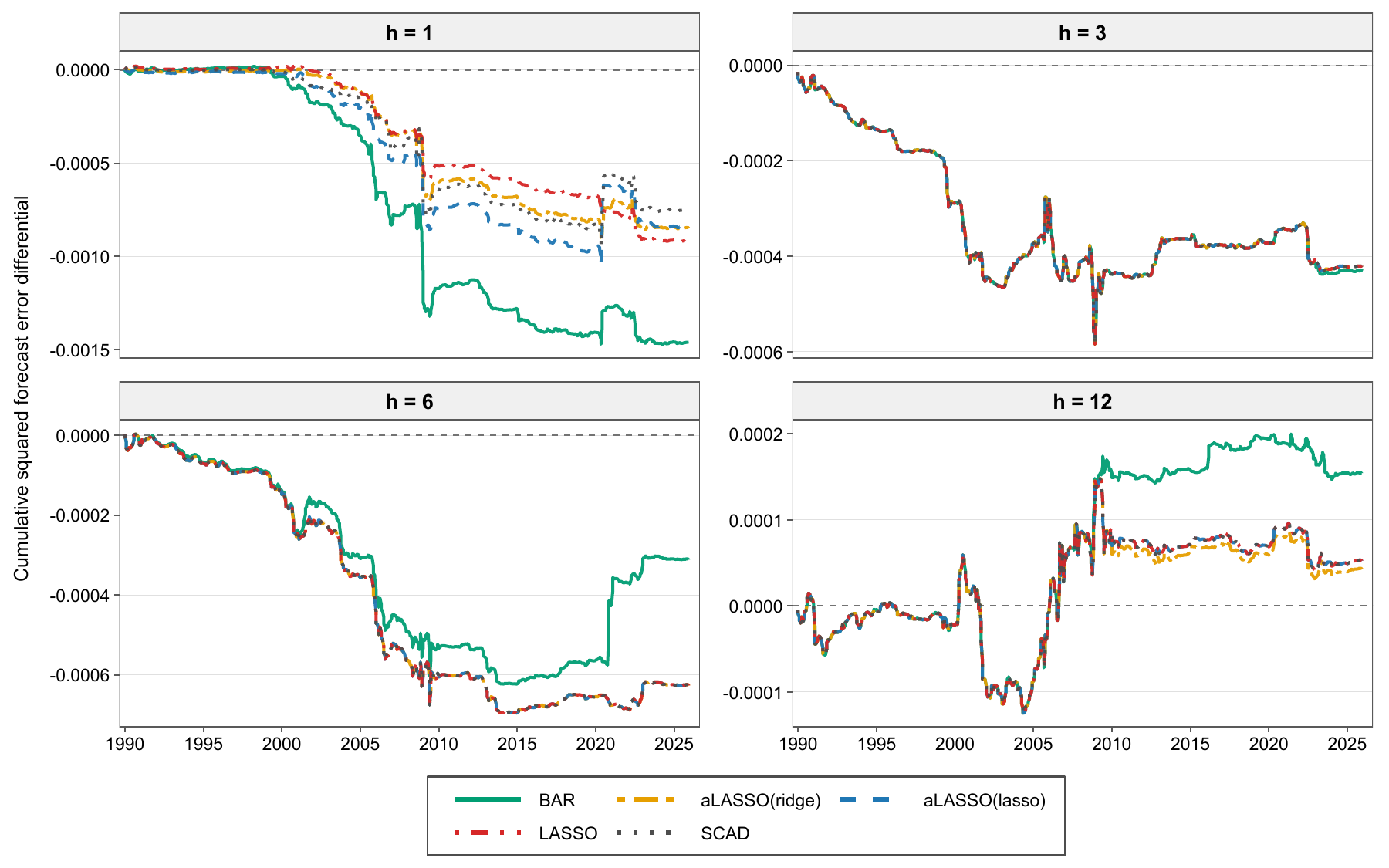}
    \label{fig:US_inflation_roll360}
\end{figure}

An insight on the evolution of forecast performance over time is provided in Figure \ref{fig:US_inflation_roll360}, which plots the cumulative sum of squared forecast error differentials (CSSFED) for each model relative to the AR(1) benchmark. A declining trajectory therefore indicates that the specific model is accumulating smaller squared forecast errors than the AR(1), whereas an increasing trajectory indicates a relative deterioration.\footnote{Since both CSSFED and RMSFE are based on squared forecast errors, periods of declining CSSFED correspond to improvements in RMSFE ratios.}
At \(h\)=1, BAR clearly outperforms the competing models: its CSSFED declines markedly between 2000 and 2010 and remains well below the other trajectories.
At \(h\)=3, the CSSFED paths are closely clustered, indicating broadly similar forecasting performance across models. By contrast, BAR deteriorates substantially at \(h\)=6 during the COVID-19 period and at \(h\)=12 in 2009, immediately after the 2008 financial crisis. Interestingly, these same episodes coincide with sizeable forecasting gains for BAR at the one-month horizon.

The results show clear horizon dependence: BAR performs best at one month, while its advantage fades at longer horizons. This is consistent with short-run inflation being driven by a more concentrated set of signals, whereas longer-horizon predictability is more diffuse and less suited to aggressive thresholding.
% ---- Plot ----
% ---- Plot ----
%%%%%%%%%%%%%%%%%%%%% CONCLUSION %%%%%%%%%%%%%%%%%%%%%
\section{Conclusion}
\noindent This paper studies the Broken Adaptive Ridge (BAR) estimator for high-dimensional VARs. We show that BAR remains theoretically well behaved under physical dependence and row-wise approximate sparsity, allowing both the dimension of the system and the number of relevant coefficients to grow with the sample size. The analysis exploits the distinctive fixed-point and thresholding structure of BAR to control the selected model and establish $\ell_1$, $\ell_2$, and prediction error rates in high-dimensional settings. Monte Carlo experiments show that BAR can achieve accurate variable selection, limited shrinkage of large coefficients, and competitive forecasting performance in high-dimensional and strongly correlated designs. In an application to U.S. inflation forecasting using FRED-MD, BAR performs particularly well at the one-month horizon, while its relative performance at longer horizons varies across economic periods. Overall, the results indicate that BAR provides a theoretically justified and practically competitive regularization method for high-dimensional time-series models.

\bibliography{literature.bib}

\newpage

% ============================================================
% APPENDIX
% ============================================================
% Numbering of equations, tables and figures
\setcounter{equation}{0}
\setcounter{table}{0}
\setcounter{figure}{0}

\renewcommand{\theequation}{\Alph{section}.\arabic{equation}}
\renewcommand{\thetable}{\Alph{section}.\arabic{table}}
\renewcommand{\thefigure}{\Alph{section}.\arabic{figure}}

% Mark the beginning of the appendix in the .toc file
\phantomsection
\addcontentsline{toc}{section}{Appendix}
\etocdepthtag.toc{appendix}

\appendix

% Appendix title
\begin{center}
    {\Large\bfseries Appendix}
\end{center}
 
% Appendix table of contents
\begingroup
\etocsettagdepth{main}{none}
\etocsettagdepth{appendix}{subsubsection}
\tableofcontents
\endgroup

\vspace{1em}

\clearpage

% ============================================================
% APPENDIX A
% ============================================================
\section{Additional Technical Details and Auxiliary Results}
\label{app:additional_technical_results}

This appendix collects supporting derivations, computational details,
and proofs omitted from the main text.

%%%%%%%%%%%%%%%%%%%%%%%%%%%%%%%%%%%%%%%%%%%%%%%%%%%%

\subsection{Additional Details on the Assumptions and VAR Setup}
\label{app:assumptions_var_details}

\subsubsection{Further Discussion of the Physical-Dependence Assumption}
\label{app:physical_dependence_details}

Assumption \ref{ass_phidep} uses the physical-dependence framework of \citet{wu2005nonlinear} to quantify temporal dependence through the data-generating mechanism itself. Assumption \ref{ass_phidep}(ii) represents $\bh_t$ as a measurable function of the innovation history $\mathcal F^t=(\ldots,\be_{t-1},\be_t)$. The coupled history $\mathcal F^{t,0}$ replaces only $\be_0$ by an independent copy, leaving every other innovation unchanged. Comparing $\bh_t=G(\mathcal F^t)$ with $\bh_{t,0}=G(\mathcal F^{t,0})$ therefore isolates the effect of a single innovation on the process $t$ periods later.

For an individual coordinate $h_{j,t}$, the functional-dependence measure $\delta_{q,j,t}=\|h_{j,t}-h_{j,t,0}\|_{L_q}$ quantifies the magnitude of this effect at horizon $t$, while \begin{align*}
\Delta_{q,j,m} = \sum_{t=m}^{\infty}\delta_{q,j,t} \end{align*}
measures its remaining cumulative effect from horizon $m$ onward. The dependence-adjusted norm \begin{align*}
\|\bh_{j,\bullet}\|_{q,\alpha} = \sup_{m\ge0}(m+1)^\alpha\Delta_{q,j,m} \end{align*} weights the cumulative effect remaining after horizon $m$ by the factor $(m+1)^\alpha$. Requiring this norm to be bounded implies $\Delta_{q,j,m} = O\bigl((m+1)^{-\alpha}\bigr)$, so that the cumulative effect decays at least polynomially. The quantity $\Psi_{q,\alpha} =\max_{1\le j\le2K}\|\bh_{j,\bullet}\|_{q,\alpha}$ extends this control uniformly over the individual coordinates of the system. In particular, $\Psi_{q,0}<\infty$ requires the effect of a single innovation to be summable over future horizons, whereas boundedness of $\Psi_{q,\alpha}$ additionally controls the rate at which the remaining effect vanishes.

Coordinatewise control, however, does not by itself control the largest perturbation across the coordinates of a growing system. This motivates \begin{align*} \omega_{q,t} = \bigl\|\|\bh_t-\bh_{t,0}\|_\infty\bigr\|_{L_q}, \qquad \Omega_{q,m} = \sum_{t=m}^{\infty}\omega_{q,t}, \end{align*} and the associated maximal dependence-adjusted norm \begin{align*} \mathcal H_{q,\alpha,T} = \sup_{m\ge0}(m+1)^\alpha\Omega_{q,m}. \end{align*}
Thus, while $\Psi_{q,\alpha}$ controls the effect of the innovation replacement one coordinate at a time, $\mathcal H_{q,\alpha,T}$ controls the largest such perturbation over all coordinates. The distinction is purely cross-sectional: both quantities use the same moment order $q$ and are based on the same single-innovation perturbation.

This additional maximal control is needed because the stochastic objects entering the BAR theory are themselves high dimensional. In fact, Lemma \ref{lem_EPB} simultaneously controls the empirical score and all entries of the sample Gram matrix. These quantities are products of coordinates of the underlying process and are collected in the proof into a single second-order process. The vector-valued maximal inequality of \citet{zhang2017gaussian}, used through the second-order construction of \citet{margaritella2025precision}, then requires control of the largest perturbation across the system. Condition \eqref{eq:linfty_dan_growth} imposes the corresponding growth restriction.

The dependence-adjusted norms are high-level quantities, but they can be related to primitive restrictions for familiar time-series models. For example, \citet[][Online Appendix C2]{chernozhukov2021lasso} provides sufficient conditions for VARMA-type and multivariate-GARCH processes, including high-dimensional VARs. 

The following simple example illustrates the innovation-replacement construction.

\begin{example}[Physical dependence in an AR(1)]
\label{ex_phidep}
Consider the stationary AR(1) \begin{align*} y_t=\phi y_{t-1}+e_t, \qquad |\phi|<1, \end{align*} where $\{e_t\}$ is iid, mean zero, and $\mathbb E|e_t|^q<\infty$. Its causal representation is \begin{align*} y_t = \sum_{\ell=0}^{\infty}\phi^\ell e_{t-\ell}. \end{align*} Let $e_0^*$ be an independent copy of $e_0$ and construct $y_{t,0}$ by replacing only $e_0$ with $e_0^*$. For $t\ge0$, \begin{align*} y_t-y_{t,0} = \phi^t(e_0-e_0^*), \end{align*} and hence \begin{align*} \delta_{q,t} &= |\phi|^t\|e_0-e_0^*\|_{L_q},\\ \Delta_{q,m} &= \frac{|\phi|^m}{1-|\phi|} \|e_0-e_0^*\|_{L_q},\\ \|y_\bullet\|_{q,\alpha} &= \sup_{m\ge0} (m+1)^\alpha\Delta_{q,m} < \infty \end{align*} for every $\alpha\ge0$, since geometric decay dominates any polynomial factor.

In the univariate case the coordinatewise and maximal measures coincide. Their distinction appears in multivariate systems. For example, for $K$ parallel AR(1) processes with the same autoregressive coefficient, \begin{align*} \omega_{q,t} = |\phi|^t \left\| \max_{1\le j\le K} |e_{j,0}-e_{j,0}^*| \right\|_{L_q}. \end{align*} The temporal decay remains geometric, while the magnitude of the largest contemporaneous perturbation may increase with $K$. The same interpretation applies to a stable VAR($p$), where the perturbation is propagated through the matrices $\bA_1,\ldots,\bA_p$.
\end{example}

\iffalse
\subsubsection{Final-Equation Representation}
\label{app:var_order_details}

The final-equation representation used in Remark \ref{rmk_laglength} follows directly from the VAR lag polynomial. Write
\begin{align*}
\bA(L) := \bI_K-\sum_{\ell=1}^p\bA_\ell L^\ell, \qquad \bA(L)\by_t=\bu_t.
\end{align*}
Let $\bA^*(L)$ denote the adjugate of $\bA(L)$. Since $\bA^*(L)\bA(L) = \det\{\bA(L)\}\bI_K$, premultiplication gives \begin{align*}
\det\{\bA(L)\}\by_t = \bA^*(L)\bu_t.
\end{align*}
The scalar polynomial $\det\{\bA(L)\}$ has degree at most $Kp$, whereas each entry of $\bA^*(L)$ has degree at most $(K-1)p$. Consequently, each component $y_{i,t}$ admits a univariate ARMA representation with orders no larger than $\operatorname{ARMA}(Kp,(K-1)p)$.  \fi
%%%%%%%%%%%%%%%%%%%%%%%%%%%%%%%%%%%%%%%%%%%%%%%%%%%%

\subsection{Computational Details of the BAR Estimator}
\label{app:bar_computational_details}

\subsubsection{Selection of the Outer-Root Branch}
\label{app:coordinate_update_derivation}

In Section \ref{sec_model_setup}, the coordinatewise BAR update selects the outer solution of the quadratic fixed-point equation whenever the activation threshold is satisfied. We now show that this choice corresponds to the locally stable branch of the underlying scalar weighted-Ridge fixed-point map.

Fix the remaining coordinates and let $b_j$ denote the corresponding partial score, with $a_j:=T^{-1}\bX_j'\bX_j$. If $z$ denotes the previous value of coordinate $j$, its weighted-Ridge update is \begin{align*} h_j(z) = \frac{b_j}{a_j+\lambda/z^2} = \frac{b_jz^2}{a_jz^2+\lambda}. \end{align*} A nonzero fixed point $z=h_j(z)$ therefore satisfies \begin{align*} a_jz^2-b_jz+\lambda=0. \end{align*} When $|b_j|>2\sqrt{\lambda a_j}$, this equation has an inner and an outer root. At either nonzero fixed point, \begin{align*} h_j'(z) = \frac{2b_jz\lambda}{(a_jz^2+\lambda)^2} = \frac{2\lambda}{a_jz^2+\lambda}, \end{align*} where the second equality follows from $b_jz=a_jz^2+\lambda$. For the outer root, $a_jz^2>\lambda$, and hence $|h_j'(z)|<1$, whereas for the inner root $a_jz^2<\lambda$ and $|h_j'(z)|>1$. Thus, conditional on the remaining coordinates, the outer root is the locally stable branch of the scalar weighted-Ridge fixed-point map. Following \citet{kawaguchi2021scalable}, Algorithm \ref{alg:BAR_CD} uses this branch and sets the coordinate to zero when $|b_j|<2\sqrt{\lambda a_j}$.

\subsubsection{Additional Implementation Details}
\label{app:implementation_details}

When $Kp\gg T$, the Ridge initializer need not be computed by inverting a $(Kp)\times(Kp)$ matrix. Indeed, \begin{align*} \hat{\bbeta}^{(0)} &= \left(T^{-1}\bX'\bX+\xi\bI\right)^{-1} T^{-1}\bX'\by\\ &= \bX' \left(\bX\bX'+T\xi\bI_T\right)^{-1} \by, \end{align*} so the inversion can instead be performed in the $T$-dimensional sample space. Alternatively, the Ridge initializer may itself be computed by coordinate descent. For partial residual $\br_j=\by-\bX_{-j}\hat{\bbeta}_{-j}$, the corresponding update is \begin{align*} \hat\beta_j = \frac{T^{-1}\bX_j'\br_j} {T^{-1}\bX_j'\bX_j+\xi}. \end{align*} Both approaches avoid forming or inverting a large dense Gram matrix. 

For the stopping rule in Algorithm \ref{alg:BAR_CD}, we use $N_{\max}=10^3$ and $\epsilon=10^{-4}$ in the simulations and applications. Increasing $N_{\max}$ or reducing $\epsilon$ further produced no material improvement in coefficient accuracy in the implementations considered here, while increasing computational cost. 

\subsubsection{Grouping Property of the BAR Fixed Point}
\label{app:grouping_property}

The following result provides the grouping property stated in Remark \ref{rmk_grouping}.
 \begin{proposition}[Grouping property]
\label{prop:grouping}
Suppose that the predictors are sample standardized so that $\|\bX_j\|_2^2=T$, and define $\hat\rho_{jk}:=T^{-1}\bX_j'\bX_k$. Let $\hat{\bbeta}_{BAR}$ be a BAR fixed point and $\hat{\br}:=\by-\bX\hat{\bbeta}_{BAR}$. Then, for any $j,k\in\hat S$, \begin{align} \left| \frac{1}{\hat\beta_{BAR,j}} - \frac{1}{\hat\beta_{BAR,k}} \right| &\le \frac{\|\hat{\br}\|_2}{\lambda\sqrt T} \sqrt{2(1-\hat\rho_{jk})} \le \frac{\|\by\|_2}{\lambda\sqrt T} \sqrt{2(1-\hat\rho_{jk})}.
\label{eq:grouping_positive}
\end{align}
Similarly, \begin{align} \left| \frac{1}{\hat\beta_{BAR,j}} + \frac{1}{\hat\beta_{BAR,k}} \right| &\le \frac{\|\hat{\br}\|_2}{\lambda\sqrt T} \sqrt{2(1+\hat\rho_{jk})} \le \frac{\|\by\|_2}{\lambda\sqrt T} \sqrt{2(1+\hat\rho_{jk})}. \label{eq:grouping_negative} \end{align}
\end{proposition}

\begin{proof}
For $j,k\in\hat S$, the fixed-point identity \eqref{eq:fixed_point_identity} gives \begin{align*} T^{-1}\bX_j'\hat{\br} = \frac{\lambda}{\hat\beta_{BAR,j}}, \qquad T^{-1}\bX_k'\hat{\br} = \frac{\lambda}{\hat\beta_{BAR,k}}. \end{align*} Hence, by Cauchy--Schwarz, \begin{align*} \lambda \left| \frac{1}{\hat\beta_{BAR,j}} - \frac{1}{\hat\beta_{BAR,k}} \right| &= T^{-1} \left| (\bX_j-\bX_k)'\hat{\br} \right|\\ &\le T^{-1} \|\bX_j-\bX_k\|_2 \|\hat{\br}\|_2\\ &= \frac{\|\hat{\br}\|_2}{\sqrt T} \sqrt{2(1-\hat\rho_{jk})}, \end{align*} where the last equality follows from $\|\bX_j\|_2^2=\|\bX_k\|_2^2=T$.  To show $\|\hat{\br}\|_2\le\|\by\|_2$, multiply the fixed-point identity by $\hat\beta_{BAR,j}$ and sum over $j\in\hat S$. Since $\hat\beta_{BAR,j}=0$ outside $\hat S$, \begin{align*} \lambda|\hat S| = T^{-1} (\bX\hat{\bbeta}_{BAR})' (\by-\bX\hat{\bbeta}_{BAR}). \end{align*} Therefore, \begin{align*} T^{-1}\|\hat{\br}\|_2^2 &= T^{-1}\|\by\|_2^2 - T^{-1}\|\bX\hat{\bbeta}_{BAR}\|_2^2 - 2\lambda|\hat S|\\ &\le T^{-1}\|\by\|_2^2. \end{align*} This proves \eqref{eq:grouping_positive}. The proof of \eqref{eq:grouping_negative} is identical after replacing $\bX_j-\bX_k$ by $\bX_j+\bX_k$, since \begin{align*} \|\bX_j+\bX_k\|_2^2 = 2T(1+\hat\rho_{jk}). \end{align*}
\end{proof}

%\newpage
% ============================================================
% APPENDIX B
% ============================================================
\section{Proofs}\label{app:proofs}

Throughout this section,
$\hat{\bSigma}_x:=T^{-1}\bX'\bX$ denotes the sample Gram matrix and
$\bSigma_x:=\mathbb E(\bx_t\bx_t')$ its population counterpart.
For the fixed-point calculations we use the standardized representation
described in Remark \ref{rmk_stand}. By Lemma \ref{lem_EPB},
$\max_j|T^{-1}\bX_j'\bX_j-1|=o_p(1)$ under population
standardization, so this normalization does not affect the rates below.

%%%%%%%%%%%%%%%%%%%%%%%%%%%%%%%%%%%%%%%%%%%%%%%%%%%%

\begin{proof}[Proof of Lemma \ref{lem_tailsize}]
Suppose first that $q_s\in(0,1)$. We use the bound $\mathbb{1}\{x>a\}\le(x/a)^{q_s}$ for $x,a>0$, the fact that $|\beta_j|\le2\sqrt{\lambda}$ for every $j\in S_\lambda^c$, and the approximate-sparsity condition $\sum_{j=1}^{Kp}|\beta_j|^{q_s}\le s_0$ from Assumption \ref{ass_sparsity}. Then, \begin{align*}
|S_\lambda| &= \sum_{j=1}^{Kp} \mathbb{1}\{|\beta_j|>2\sqrt{\lambda}\} \le (2\sqrt{\lambda})^{-q_s} \sum_{j=1}^{Kp}|\beta_j|^{q_s} \le 2^{-q_s}s_0\lambda^{-q_s/2}, \\ \|\bbeta_{S_\lambda^c}\|_r^r &= \sum_{j\in S_\lambda^c} |\beta_j|^{r-q_s}|\beta_j|^{q_s} \le (2\sqrt{\lambda})^{r-q_s} \sum_{j\in S_\lambda^c}|\beta_j|^{q_s} \le s_0(2\sqrt{\lambda})^{r-q_s}, \qquad r\in\{1,2\}. \end{align*}
Taking the $r$th root in the second inequality gives $\|\bbeta_{S_\lambda^c}\|_r \lesssim s_0^{1/r}\lambda^{1/2-q_s/(2r)}$.

When $q_s=0$, Assumption \ref{ass_sparsity} is exact sparsity and $\|\bbeta\|_0\le s_0$. Hence, \begin{align*} |S_\lambda| &\le s_0, & \|\bbeta_{S_\lambda^c}\|_r^r &\le s_0(2\sqrt{\lambda})^r, & \|\bbeta_{S_\lambda^c}\|_r &\lesssim s_0^{1/r}\sqrt{\lambda}, \end{align*} which yields the same bounds at $q_s=0$.
\end{proof}

%%%%%%%%%%%%%%%%%%%%%%%%%%%%%%%%%%%%%%%%%%%%%%%%%%%%

\begin{proof}[Proof of Lemma \ref{lem_joint_concentration}]
Let $d:=Kp$, $r:=q/2>2$, $\ell_T:=2\log(d+1)$, and define \begin{align*}
\bg_t &:=(\bx_t',u_t)'\in\mathbb R^{d+1}, & \bchi_t &:= \operatorname{vec} \left\{ \bg_t\bg_t' - \mathbb E(\bg_t\bg_t') \right\}, & \bar{\bchi}_T &:= T^{-1}\sum_{t=1}^T\bchi_t. \end{align*}
Let $\bg_{t,0}$ and $\bchi_{t,0}$ denote the corresponding coupled processes. We further define \begin{align*} \Psi_{q,\alpha}^{(g)} &:= \max_{1\le j\le d+1} \|g_{j,\bullet}\|_{q,\alpha}, & \mathcal G_{q,\alpha,T} &:= \sup_{m\ge0}(m+1)^\alpha \sum_{t=m}^\infty \bigl\| \|\bg_t-\bg_{t,0}\|_\infty \bigr\|_{L_q}.
\end{align*}
Since $p$ is fixed and the components of $\bg_t$ are either coordinates of $\bu_t$ or fixed lags of coordinates of $\by_t$, Assumption \ref{ass_phidep} implies \begin{align*} \Psi_{q,0}^{(g)} &\lesssim\Psi_{q,0}, & \Psi_{q,\alpha}^{(g)} &\lesssim\Psi_{q,\alpha}, & \mathcal G_{q,0,T} &\lesssim\mathcal H_{q,0,T}, & \mathcal G_{q,\alpha,T} &\lesssim\mathcal H_{q,\alpha,T}. \end{align*}

Furthermore, a generic coordinate of $\bchi_t$, indexed by $a=(j,k)$, satisfies, by Minkowski's and H\"older's inequalities, \begin{align*} \|\chi_{a,t}-\chi_{a,t,0}\|_{L_r} &\le \|(g_{j,t}-g_{j,t,0})g_{k,t}\|_{L_r} + \|g_{j,t,0}(g_{k,t}-g_{k,t,0})\|_{L_r}\\ &\le \|g_{j,t}-g_{j,t,0}\|_{L_q}\|g_{k,t}\|_{L_q} + \|g_{j,t,0}\|_{L_q} \|g_{k,t}-g_{k,t,0}\|_{L_q}. \end{align*} The standard product inequality for dependence-adjusted norms \citep[Section 3.1]{zhang2017gaussian} therefore gives \begin{align} A_{2,T} &:= \max_{1\le a\le(d+1)^2} \|\chi_{a,\bullet}\|_{2,\alpha} \lesssim \Psi_{4,0}^{(g)} \Psi_{4,\alpha}^{(g)} \lesssim \Psi_{q,0}\Psi_{q,\alpha}, \notag\\ B_{r,T} &:= \sup_{m\ge0}(m+1)^\alpha \sum_{t=m}^\infty \bigl\| \|\bchi_t-\bchi_{t,0}\|_\infty \bigr\|_{L_r} \lesssim \mathcal G_{q,0,T} \mathcal G_{q,\alpha,T} \lesssim \mathcal H_{q,0,T} \mathcal H_{q,\alpha,T}. \label{app:eq:chi_norm_inheritance} \end{align} By \eqref{eq:coordinate_dan_bound}, $A_{2,T}=O(1)$. Indeed, since $q>4$, the $L^4$ dependence norms in the first line are controlled by their $L^q$ counterparts. Moreover, \eqref{eq:linfty_dan_growth} implies \begin{align} B_{r,T}\ell_T = o\left( T^{1/2-\alpha_1/r} \right). \label{app:eq:B_growth} \end{align}

Following the centered second-order construction in \citet{margaritella2025precision}, we apply the Fuk-Nagaev bound of Theorem 6.2 of \citet{zhang2017gaussian} with \begin{align} \kappa_{T,d} := C_0 \left\{ T^{-1/2}\ell_T^{1/2}A_{2,T} + T^{\alpha_1/r-1}\ell_T^{3/2}B_{r,T} \right\}. \label{app:eq:kappa_joint} \end{align} For a sufficiently large fixed $C_0$ and for positive constants, denoted generically by $C_{r,\alpha}$, depending only on $(r,\alpha)$, \begin{align} \mathbb P \left( \|\bar{\bchi}_T\|_\infty>\kappa_{T,d} \right) &\le C_{r,\alpha} \frac{ T^{\alpha_1-r}\ell_T^{r/2}B_{r,T}^r }{ \kappa_{T,d}^r } + C_{r,\alpha} \exp \left\{ - \frac{ C_{r,\alpha}T\kappa_{T,d}^2 }{ A_{2,T}^2 } \right\}. \label{app:eq:joint_tail_bound} \end{align} By \eqref{app:eq:B_growth} and $A_{2,T}=O(1)$, \begin{align*} T^{\alpha_1/r-1}\ell_T^{3/2}B_{r,T} &= T^{-1/2}\ell_T^{1/2} \left\{ T^{\alpha_1/r-1/2}\ell_T B_{r,T} \right\} = o\left(T^{-1/2}\ell_T^{1/2}\right), \\ \kappa_{T,d} &= O\left\{ \sqrt{\frac{\log(d)}{T}} \right\}, \\ \frac{ T^{\alpha_1-r}\ell_T^{r/2}B_{r,T}^r }{ \kappa_{T,d}^r } &\lesssim \ell_T^{-r} =o(1), \qquad \frac{T\kappa_{T,d}^2}{A_{2,T}^2} \gtrsim \ell_T\to\infty. \end{align*} Both terms on the right-hand side of \eqref{app:eq:joint_tail_bound} therefore vanish. Hence, for a sufficiently large fixed $C_\chi$, \begin{align*} \mathbb P \left\{ \|\bar{\bchi}_T\|_\infty \le C_\chi \sqrt{\frac{\log(d)}{T}} \right\} \to1. \end{align*} Finally, by construction, the coordinates of $T^{-1}\bX'\bu$ and $\operatorname{vec}(\hat{\bSigma}_x-\bSigma_x)$ are contained in $\bar{\bchi}_T$. In particular, $\mathbb E(\bx_tu_t)=0$ identifies the score block, while the $\bx_t\bx_t'$ block yields the centered sample Gram matrix. Therefore, with $d=Kp$, \begin{align*} \left\|T^{-1}\bX'\bu\right\|_\infty \vee \left\|\hat{\bSigma}_x-\bSigma_x\right\|_{\max} = O_p\left( \sqrt{\frac{\log(Kp)}{T}} \right), \end{align*} which proves Lemma \ref{lem_joint_concentration}.
\end{proof}

%%%%%%%%%%%%%%%%%%%%%%%%%%%%%%%%%%%%%%%%%%%%%%%%%%%%

\begin{proof}[Proof of Lemma \ref{lem_BARshell}]
For every $j\in U^c$, $j\notin\hat S$ and therefore $\hat\beta_{BAR,j}=0$. Hence, \begin{align*} \bDelta_{U^c} &= -\bbeta_{U^c}, & \bDelta &= \bDelta_U-\bbeta_{U^c}, & \|\bDelta_{U^c}\|_r &= \|\bbeta_{U^c}\|_r \le \|\bbeta_{S_\lambda^c}\|_r, \qquad r\in\{1,2\}. \end{align*} The final bound follows by the triangle inequality and Corollary \ref{cor_rates}. \end{proof}

%%%%%%%%%%%%%%%%%%%%%%%%%%%%%%%%%%%%%%%%%%%%%%%%%%%%
\begin{proof}[Proof of Proposition \ref{prop_crude_bound}]
For every $j\in\hat S$, \eqref{eq:fixed_point_identity} gives $v_j=\lambda/\hat\beta_{BAR,j}$. Therefore, \begin{align*} \lambda|\hat S| &= \sum_{j\in\hat S}\hat\beta_{BAR,j}v_j = T^{-1} (\bX\hat{\bbeta}_{BAR})' (\by-\bX\hat{\bbeta}_{BAR}) \\ &= \frac{1}{4T}\|\by\|_2^2 - \frac{1}{T} \left\| \bX\hat{\bbeta}_{BAR} -\frac{\by}{2} \right\|_2^2 \le \frac{1}{4T}\|\by\|_2^2, \end{align*} where we used $\hat\beta_{BAR,j}=0$ for $j\notin\hat S$. Consequently, \begin{align*} |\hat S| &\le \frac{\|\by\|_2^2}{4\lambda T}, \\ |U| &\le |S_\lambda|+|\hat S| \le 2^{-q_s}s_0\lambda^{-q_s/2} + \frac{\|\by\|_2^2}{4\lambda T} \le \bar m_T \qquad\text{on }\mathcal E_y, \end{align*} where the second line uses Lemma \ref{lem_tailsize} and the definition of $\mathcal E_y$. \end{proof}

%%%%%%%%%%%%%%%%%%%%%%%%%%%%%%%%%%%%%%%%%%%%%%%%%%%%

\begin{proof}[Proof of Lemma \ref{lem_shellident}]
Using $\by=\bX\bbeta+\bu$ and $\bDelta=\hat{\bbeta}_{BAR}-\bbeta$, \begin{align*} \bv &= T^{-1}\bX' (\by-\bX\hat{\bbeta}_{BAR}) = T^{-1}\bX'\bu-\hat{\bSigma}_x\bDelta, \\ \bv_U &= T^{-1}\bX_U'\bu - \hat{\bSigma}_{x,UU}\bDelta_U - \hat{\bSigma}_{x,UU^c}\bDelta_{U^c} \\ &= T^{-1}\bX_U'\bu - \hat{\bSigma}_{x,UU}\bDelta_U + \hat{\bSigma}_{x,UU^c}\bbeta_{U^c}, \end{align*}
where the last equality follows from Lemma \ref{lem_BARshell}. Premultiplication by $\bDelta_U'$ and rearrangement give \eqref{eq_shell_ident}. \end{proof}

%%%%%%%%%%%%%%%%%%%%%%%%%%%%%%%%%%%%%%%%%%%%%%%%%%%%

\begin{proof}[Proof of Lemma \ref{lem_false_positives}]
For $q_s\in(0,1)$, \begin{align*} |F_B| &\le |B_\lambda| \le \sum_{j=1}^{Kp} \mathbb{1}\{|\beta_j|>c_0\sqrt{\lambda}\} \\ &\le (c_0\sqrt{\lambda})^{-q_s} \sum_{j=1}^{Kp}|\beta_j|^{q_s} \le c_0^{-q_s}s_0\lambda^{-q_s/2} \lesssim s_0\lambda^{-q_s/2}. \end{align*} When $q_s=0$, every coordinate in $B_\lambda$ is nonzero and exact sparsity gives $|F_B|\le|B_\lambda|\le\|\bbeta\|_0\le s_0$, which is the same bound at $q_s=0$. \end{proof} 

%%%%%%%%%%%%%%%%%%%%%%%%%%%%%%%%%%%%%%%%%%%%%%%%%%%%

\begin{proof}[Proof of Lemma \ref{lem_deep_tail_penalty}]
For every $j\in F_D$, $j\in\hat S$ and therefore, by \eqref{eq:fixed_point_identity}, \begin{align*} \Delta_jv_j &= (\hat\beta_{BAR,j}-\beta_j) \frac{\lambda}{\hat\beta_{BAR,j}} = \lambda \left( 1-\frac{\beta_j}{\hat\beta_{BAR,j}} \right) \\ &\ge \lambda \left( 1- \left| \frac{\beta_j}{\hat\beta_{BAR,j}} \right| \right) \ge (1-c_0)\lambda, \end{align*} where the last inequality uses $|\beta_j|\le c_0\sqrt{\lambda}$ from equation \ref{eq:buffer_deep_tail} in the main text and $|\hat\beta_{BAR,j}|\ge\sqrt{\lambda}$. Summing over $F_D$ yields \begin{align*} \sum_{j\in F_D}\Delta_jv_j \ge (1-c_0)\lambda|F_D|. \end{align*} \end{proof}

%%%%%%%%%%%%%%%%%%%%%%%%%%%%%%%%%%%%%%%%%%%%%%%%%%%%

\begin{proof}[Proof of Proposition \ref{prop_shell_dimension}]
Let \begin{align*}
J:=U\setminus F_D=S_\lambda\cup F_B, \qquad |J|=|S_\lambda|+|F_B|. \end{align*} By Lemmas \ref{lem_shellident} and \ref{lem_deep_tail_penalty}, \begin{align} \bDelta_U'\hat{\bSigma}_{x,UU}\bDelta_U + (1-c_0)\lambda|F_D| &\le \left| \bDelta_U'T^{-1}\bX_U'\bu \right| + \left| \bDelta_U'\hat{\bSigma}_{x,UU^c}\bbeta_{U^c} \right| + \left| \sum_{j\in J}\Delta_jv_j \right|. \label{eq_shell_master_prop} \end{align}

On $\mathcal E_\chi$, Lemma \ref{lem_EPB} gives
$\|T^{-1}\bX'\bu\|_\infty\le c_u\sqrt{\lambda}$, where
$c_u:=C_\chi/\sqrt{C_\lambda}$. Moreover, $|v_j|\le2\sqrt{\lambda}$ for all $j\in J$: if $j\in J\cap\hat S$, this follows from $v_j=\lambda/\hat\beta_{BAR,j}$ and $|\hat\beta_{BAR,j}|\ge\sqrt{\lambda}$; if $j\in J\cap\hat S^c$, then $j\in S_\lambda\cap\hat S^c$ and inactivity implies $|v_j|=|b_j|<2\sqrt{\lambda}$. Hence, \begin{align*} & \left| \bDelta_U'T^{-1}\bX_U'\bu \right| + \left| \sum_{j\in J}\Delta_jv_j \right| \\ &\qquad\le \sqrt{\lambda} \left[ (c_u+2)\|\bDelta_J\|_1 + c_u\|\bDelta_{F_D}\|_1 \right] \\ &\qquad\le \sqrt{ (c_u+2)^2|J|\lambda + c_u^2|F_D|\lambda } \, \|\bDelta_U\|_2. \end{align*} The last step follows from Cauchy--Schwarz.  For the remaining cross-block term, \begin{align} \left| \bDelta_U' \hat{\bSigma}_{x,UU^c} \bbeta_{U^c} \right| &= \left| T^{-1} (\bX_U\bDelta_U)' (\bX_{U^c}\bbeta_{U^c}) \right| \notag\\ &\le \left( \bDelta_U'\hat{\bSigma}_{x,UU}\bDelta_U \right)^{1/2} \left( \bbeta_{U^c}' \hat{\bSigma}_{x,U^cU^c} \bbeta_{U^c} \right)^{1/2} \notag\\ &\le \frac12 \bDelta_U'\hat{\bSigma}_{x,UU}\bDelta_U + \frac12 \bbeta_{U^c}' \hat{\bSigma}_{x,U^cU^c} \bbeta_{U^c}. \label{eq_crossterm_amgm}
\end{align} Substituting the preceding bounds into \eqref{eq_shell_master_prop}, using $\bDelta_U'\hat{\bSigma}_{x,UU}\bDelta_U \ge \kappa_{\mathrm{crude}}\|\bDelta_U\|_2^2$ on $\mathcal G$, and applying Young's inequality gives \begin{align*} \frac{\kappa_{\mathrm{crude}}}{2}\|\bDelta_U\|_2^2 + (1-c_0)\lambda|F_D| &\le \sqrt{ (c_u+2)^2|J|\lambda + c_u^2|F_D|\lambda } \, \|\bDelta_U\|_2 + \frac12 \bbeta_{U^c}'\hat{\bSigma}_{x,U^cU^c}\bbeta_{U^c} \\ &\le \frac{\kappa_{\mathrm{crude}}}{4}\|\bDelta_U\|_2^2 + \frac{(c_u+2)^2}{\kappa_{\mathrm{crude}}}|J|\lambda + \frac{c_u^2}{\kappa_{\mathrm{crude}}}|F_D|\lambda + \frac12 \bbeta_{U^c}'\hat{\bSigma}_{x,U^cU^c}\bbeta_{U^c}. \end{align*} Therefore, with $a_u:=c_u^2/\kappa_{\mathrm{crude}}$, \begin{align} (1-c_0-a_u)\lambda|F_D| &\le \frac{(c_u+2)^2}{\kappa_{\mathrm{crude}}} \lambda \bigl( |S_\lambda|+|F_B| \bigr) + \frac12 \bbeta_{U^c}' \hat{\bSigma}_{x,U^cU^c} \bbeta_{U^c}. \label{eq_FD_rearranged} \end{align}

To control the last term, let $\delta_T:=\|\hat{\bSigma}_x-\bSigma_x\|_{\max}$. Since $U^c\subseteq S_\lambda^c$, \begin{align} \bbeta_{U^c}' \hat{\bSigma}_{x,U^cU^c} \bbeta_{U^c} &\le \bbeta_{U^c}' \bSigma_{x,U^cU^c} \bbeta_{U^c} + \delta_T\|\bbeta_{U^c}\|_1^2 \notag\\ &\le C_{\max} \|\bbeta_{S_\lambda^c}\|_2^2 + C_\chi \sqrt{\frac{\log(Kp)}{T}} \, \|\bbeta_{S_\lambda^c}\|_1^2 \notag\\ &\lesssim \|\bbeta_{S_\lambda^c}\|_2^2 + \sqrt{\lambda} \|\bbeta_{S_\lambda^c}\|_1^2, \label{app:eq:random_tail_pathwise} \end{align} where the second line uses $\lambda_{\max}(\bSigma_x)\le C_{\max}$ and the Gram-matrix deviation bound in Lemma \ref{lem_EPB} on $\mathcal E_\chi$, while the last line uses $\lambda=C_\lambda\log(Kp)/T$. Since $1-c_0-a_u>0$, \eqref{eq_FD_rearranged} therefore yields \begin{align*} |F_D| &\le C_1\bigl(|S_\lambda|+|F_B|\bigr) + C_2\lambda^{-1} \left\{ \|\bbeta_{S_\lambda^c}\|_2^2 + \sqrt{\lambda} \|\bbeta_{S_\lambda^c}\|_1^2 \right\}, \end{align*} which proves \eqref{eq:shell_dimension_refinement}.

Finally, Lemmas \ref{lem_tailsize} and \ref{lem_false_positives} and Corollary \ref{cor_rates} imply \begin{align*} |S_\lambda|+|F_B| &\lesssim s_0\lambda^{-q_s/2}, \\ \lambda^{-1} \|\bbeta_{S_\lambda^c}\|_2^2 &\lesssim s_0\lambda^{-q_s/2}, \\ \lambda^{-1/2} \|\bbeta_{S_\lambda^c}\|_1^2 &\lesssim s_0^2\lambda^{1/2-q_s} = \left( s_0\lambda^{-q_s/2} \right) \left( s_0\lambda^{(1-q_s)/2} \right) = o\left( s_0\lambda^{-q_s/2} \right). \end{align*} Thus, for all sufficiently large $T$, there exist fixed constants $C_D,C_U>0$ such that, on $\mathcal G$, \begin{align*} |F_D| &\le C_Ds_0\lambda^{-q_s/2}, & |U| &\le C_Us_0\lambda^{-q_s/2}. \end{align*} Since $\mathbb P(\mathcal G)\to1$, the final $\mathcal O_p(s_0\lambda^{-q_s/2})$ statement follows. \end{proof}

%%%%%%%%%%%%%%%%%%%%%%%%%%%%%%%%%%%%%%%%%%%%%%%%%%%%

\begin{proof}[Proof of Lemma \ref{lem_refined_RE}]
On $\{|U|\le m_T^\sharp\}\cap\mathcal E_\chi$, Weyl's inequality
and Lemma \ref{lem_EPB} give \begin{align*} \lambda_{\min}(\hat{\bSigma}_{x,UU}) &\ge \lambda_{\min}(\bSigma_{x,UU}) - \| \hat{\bSigma}_{x,UU} - \bSigma_{x,UU} \|_{\mathrm{op}} \\ &\ge \kappa - |U| \|\hat{\bSigma}_x-\bSigma_x\|_{\max} \\ &\ge \kappa - C_\chi m_T^\sharp \sqrt{\frac{\log(Kp)}{T}} \\ &= \kappa - O\left\{ s_0\lambda^{(1-q_s)/2} + \sqrt{\lambda} \right\} = \kappa-o(1). \end{align*} The $\sqrt{\lambda}$ term in the penultimate line comes from the ceiling in $m_T^\sharp=\lceil C_Us_0\lambda^{-q_s/2}\rceil$. Since $s_0\lambda^{(1-q_s)/2}=o(1)$ and $\lambda\to0$, the final remainder is smaller than $\kappa/2$ for all sufficiently large $T$, which proves $\lambda_{\min}(\hat{\bSigma}_{x,UU})\ge\kappa/2$. \end{proof}

%%%%%%%%%%%%%%%%%%%%%%%%%%%%%%%%%%%%%%%%%%%%%%%%%%%%

\begin{proof}[Proof of Theorem \ref{thm_estimation_rates}]
Let $J:=U\setminus F_D$. Starting from \eqref{eq_shell_ident}, using Lemma \ref{lem_deep_tail_penalty}, the bounds from the proof of Proposition \ref{prop_shell_dimension}, and \eqref{eq_crossterm_amgm}, we obtain on $\mathcal E_T$ \begin{align*} \frac12 \bDelta_U' \hat{\bSigma}_{x,UU} \bDelta_U &\le (c_u+2) \sqrt{|U|\lambda} \|\bDelta_U\|_2 + \frac12 \bbeta_{U^c}' \hat{\bSigma}_{x,U^cU^c} \bbeta_{U^c} \\ &\le (c_u+2) \sqrt{|U|\lambda} \|\bDelta_U\|_2 + O_p\left( s_0\lambda^{1-q_s/2} \right). \end{align*} Here, the last line follows from \eqref{app:eq:random_tail_pathwise}, Corollary \ref{cor_rates}, and $s_0\lambda^{(1-q_s)/2}=o(1)$. By Lemma \ref{lem_refined_RE} and Young's inequality, \begin{align} \frac{\kappa}{4}\|\bDelta_U\|_2^2 &\le (c_u+2) \sqrt{|U|\lambda} \|\bDelta_U\|_2 + \frac12 \bbeta_{U^c}' \hat{\bSigma}_{x,U^cU^c} \bbeta_{U^c} \notag\\ &\le \frac{\kappa}{8}\|\bDelta_U\|_2^2 + \frac{2(c_u+2)^2}{\kappa}|U|\lambda + \frac12 \bbeta_{U^c}' \hat{\bSigma}_{x,U^cU^c} \bbeta_{U^c}, \notag\\ \|\bDelta_U\|_2^2 &\le \frac{16(c_u+2)^2}{\kappa^2}|U|\lambda + \frac{4}{\kappa} \bbeta_{U^c}' \hat{\bSigma}_{x,U^cU^c} \bbeta_{U^c} = O_p\left( s_0\lambda^{1-q_s/2} \right), \label{eq_l2_shell_bound_compact} \end{align} where Proposition \ref{prop_shell_dimension} gives $|U|\lambda=O_p(s_0\lambda^{1-q_s/2})$. Thus \begin{align*} \|\bDelta_U\|_2 = O_p\left( \sqrt{s_0\lambda^{1-q_s/2}} \right). \end{align*}  Using Lemma \ref{lem_BARshell} and Corollary \ref{cor_rates}, \begin{align*} \|\hat{\bbeta}_{BAR}-\bbeta\|_2 &\le \|\bDelta_U\|_2 + \|\bbeta_{S_\lambda^c}\|_2 = O_p\left( \sqrt{s_0\lambda^{1-q_s/2}} \right), \\ \|\hat{\bbeta}_{BAR}-\bbeta\|_1 &\le \sqrt{|U|}\|\bDelta_U\|_2 + \|\bbeta_{S_\lambda^c}\|_1 \\ &= O_p\left( \sqrt{s_0\lambda^{-q_s/2}} \sqrt{s_0\lambda^{1-q_s/2}} \right) + O\left( s_0\lambda^{(1-q_s)/2} \right) \\ &= O_p\left( s_0\lambda^{(1-q_s)/2} \right). \end{align*}  Finally, \begin{align*} \frac1T\|\bX\bDelta\|_2^2 &= \bDelta' \hat{\bSigma}_x \bDelta \\ &= \bDelta_U' \hat{\bSigma}_{x,UU} \bDelta_U - 2\bDelta_U' \hat{\bSigma}_{x,UU^c} \bbeta_{U^c} + \bbeta_{U^c}' \hat{\bSigma}_{x,U^cU^c} \bbeta_{U^c} \\ &\le 2\bDelta_U' \hat{\bSigma}_{x,UU} \bDelta_U + 2\bbeta_{U^c}' \hat{\bSigma}_{x,U^cU^c} \bbeta_{U^c}. \end{align*} Moreover, on $\{|U|\le m_T^\sharp\}\cap\mathcal E_\chi$, \begin{align*} \lambda_{\max}(\hat{\bSigma}_{x,UU}) &\le \lambda_{\max}(\bSigma_{x,UU}) + \| \hat{\bSigma}_{x,UU} - \bSigma_{x,UU} \|_{\mathrm{op}} \\ &\le C_{\max}+o(1), \end{align*} so that \begin{align*} \bDelta_U' \hat{\bSigma}_{x,UU} \bDelta_U &= O_p\left( \|\bDelta_U\|_2^2 \right) = O_p\left( s_0\lambda^{1-q_s/2} \right), \\ \bbeta_{U^c}' \hat{\bSigma}_{x,U^cU^c} \bbeta_{U^c} &= O_p\left( s_0\lambda^{1-q_s/2} \right). \end{align*} Therefore, \begin{align*} T^{-1} \| \bX(\hat{\bbeta}_{BAR}-\bbeta) \|_2^2 = O_p\left( s_0\lambda^{1-q_s/2} \right), \end{align*} which completes the proof. \end{proof}

%%%%%%%%%%%%%%%%%%%%%%%%%%%%%%%%%%%%%%%%%%%%%%%%%%%%

\begin{proof}[Proof of Corollary \ref{cor_asymp}]
Under $Kp=O(T^\nu)$, $\log(Kp)=O(\log T)$ and $\lambda\asymp\log(Kp)/T\to0$. Hence, \begin{align*} s_0\lambda^{(1-q_s)/2} &\to0, \\ s_0\lambda^{1-q_s/2} &= \left\{ s_0\lambda^{(1-q_s)/2} \right\} \lambda^{1/2} \to0. \end{align*} The first line makes the $\ell_1$ rate vanish, while the second makes the prediction error and squared $\ell_2$ rate vanish. Bounded $s_0$ is sufficient since $q_s<1$ and $\lambda\to0$. \end{proof}

% ============================================================
% APPENDIX C
% ============================================================
\section{Additional Simulation Results}\label{app:sim}

The following tables report the detailed Monte Carlo results underlying the aggregate win rates presented in the main text.\footnote{Bold and underlined values denote the best-performing model for a given combination of $K$ and $T$. Ties are shown in bold without underlying, while no highlighting is used when no model dominates.} Whereas the win-rate table summarizes performance across values of \(K\) and \(r\), the appendix tables report results for each individual simulation design, allowing a more granular comparison of the competing estimators across sample sizes, dimensionality, sparsity levels, and innovation dependence.
In addition to the measures reported in the win-rate table, we include three further evaluation criteria:
\begin{itemize}
    \item \textbf{False positive rate}: fraction of truly irrelevant coefficients that are incorrectly estimated as non-zero,
    \begin{align*}
    \mathrm{False \ positive \ rate}
    = \frac{\#\bigl\{ j : \widehat{\beta}_j \neq 0 \wedge \beta_j = 0 \bigr\}}
           {\#\bigl\{ j : \beta_j = 0 \bigr\}},
    \end{align*}
    capturing the risk of including irrelevant variables.
    \item \textbf{False negative rate}: fraction of truly relevant coefficients that are incorrectly estimated as zero,
    \begin{align*}
    \mathrm{False \ negative \ rate}
    = \frac{\#\bigl\{ j : \widehat{\beta}_j = 0 \wedge \beta_j \neq 0 \bigr\}}
           {\#\bigl\{ j : \beta_j \neq 0 \bigr\}},
    \end{align*}
    capturing the risk of excluding relevant variables. 
    \item \textbf{Number of variables}: the number of variables selected by the model, including both truly relevant and irrelevant variables.
\end{itemize}

\foreach \corr/\criteria/\r in {
    0/ebic/1,
    0.5/ebic/1,
    0/ebic/2,
    0.5/ebic/2,
    0/ebic/5,
    0.5/ebic/5,
    0/ebic/10,
    0.5/ebic/10%
    }{\input{Tables/tab_MC_BAR_\criteria_CMP_rho\corr_sparsity_r\r_eig_0.75_0.25}
    }

To provide a more detailed assessment of the support recovery, we complement the Monte Carlo tables with heatmaps of the estimated coefficient matrices.
Each cell reports the element-wise mean absolute estimate,
\begin{align*}
\frac{1}{n}
\sum_{m=1}^{n}
\left\lvert \widehat{\Phi}_{ij}^{(m)} \right\rvert,
\end{align*}
where \(\widehat{\Phi}_{ij}^{(m)}\) denotes the coefficient in position \((i,j)\) estimated in the \(m^{th}\) Monte Carlo simulation. Taking absolute values prevents coefficients with opposite signs from cancelling across simulations and therefore provides a graphical representation of their average estimated magnitude.
Darker blue cells indicate larger average coefficient magnitudes, whereas lighter blue cells correspond to coefficients estimated to be close to zero.
Because the heatmap entries are averages across MC simulations, coefficients selected only occasionally may still produce small positive entries. To avoid visually overstating such negligible estimates, mean absolute coefficients below \(10^{-3}\) are treated as zero and displayed in white; without this threshold, the heatmaps would appear almost entirely colored.
For readability, we report the designs with \(K\)=20, \(T\)=500, \(\rho\)=0, and \(r\)\(\in\)\{1,2,5,10\}, with the regularization parameter \(\lambda\) selected by the EBIC.

\foreach \K/\T/\criteria/\r in {
    20/500/ebic/1,
    20/500/ebic/2,
    20/500/ebic/5,
    20/500/ebic/10%
    }{
    \begin{figure}[H] 
      \centering
        \caption{Heatmap of $|\hat{\Phi}|$ for $r = \r$.}
      \includegraphics[width=0.8\textwidth]
      {Plots/heatmap_expCMP_\criteria_r\r_K\K_T\T_rho0.pdf}
      \label{fig:heatmap_ebic_r\r_K\K_T\T}
    \end{figure}
    }

% ============================================================
% APPENDIX D
% ============================================================
\section{Additional Empirical Results}

This section assesses the robustness of the empirical findings to the estimation window scheme. We retain the same dataset, forecasting equation, forecast horizons, standardization procedure, AR(1) benchmark, and evaluation measures used in the main empirical analysis.
We then compare the baseline rolling window results with those obtained using an expanding window, thereby assessing whether the strong short-horizon performance of BAR, as well as its more heterogeneous performance at longer horizons, is sensitive to the estimation-window scheme. The results are fully confirmed for the pre-COVID period, for which the two estimation schemes yield identical conclusions, and remain broadly consistent for the COVID-19 and post-pandemic periods, despite some minor differences in relative performance.

\foreach /\criteria/\roll in {
    ebic/0%
    }{\input{Tables/tab_BAR_MD_\criteria_CPIAUCSL_lags1_roll\roll}
    }

\end{document}